\documentclass[twocolumn]{aastex631}

\usepackage{newtxtext,newtxmath}
\usepackage{orcidlink}

\usepackage{booktabs}
\usepackage{array}

\usepackage{natbib}
\defcitealias{2018ApJ...857..121Y}{Yuan18}

\begin{document}

\title{A systematic study of AGN feedback in a disk galaxy I: global overview}

\correspondingauthor{Feng Yuan, Suoqing Ji}
\email{fyuan@fudan.edu.cn, sqji@fudan.edu.cn}

\author{Yuxuan Zou\orcidlink{0009-0006-4662-3053}}
\affiliation{Astrophysics Division, Shanghai Astronomical Observatory, Chinese Academy of Sciences, 80 Nandan Road, Shanghai 200030, China}
\affiliation{University of Chinese Academy of Sciences, No. 19A Yuquan Road, Beijing 100049, China}

\author{Feng Yuan\orcidlink{0000-0003-3564-6437}}
\affiliation{Center for Astronomy and Astrophysics and Department of Physics, Fudan University, Shanghai 200438, China}

\author{Suoqing Ji\orcidlink{0000-0001-9658-0588}}
\affiliation{Center for Astronomy and Astrophysics and Department of Physics, Fudan University, Shanghai 200438, China}
\affiliation{Key Laboratory of Nuclear Physics and Ion-Beam Application (MOE), Fudan University, Shanghai 200433, China}

\author{Luis C. Ho\orcidlink{0000-0001-6947-5846}}
\affiliation{Kavli Institute for Astronomy and Astrophysics, Peking University, Beijing 100871, China}
\affiliation{Department of Astronomy, School of Physics, Peking University, Beijing 100871, China}

\author{Yingjie Peng\orcidlink{0000-0003-0939-9671}}
\affiliation{Kavli Institute for Astronomy and Astrophysics, Peking University, Beijing 100871, China}
\affiliation{Department of Astronomy, School of Physics, Peking University, Beijing 100871, China}

\author{Jing Wang\orcidlink{0000-0002-6593-8820}}
\affiliation{Kavli Institute for Astronomy and Astrophysics, Peking University, Beijing 100871, China}

\author{Bocheng Zhu\orcidlink{0000-0003-0900-4481}}
\affiliation{National Astronomical Observatories, Chinese Academy of Sciences, 20A Datun Road, Beijing 100101, China}

\author{Tao Wang\orcidlink{0000-0002-2504-2421}}
\affiliation{School of Astronomy and Space Science, Nanjing University, Nanjing 210093, China}
\affiliation{Key Laboratory of Modern Astronomy and Astrophysics, Nanjing University, Ministry of Education, Nanjing 210093, China}

\begin{abstract}
This is the first paper in a series using our MACER framework to investigate the evolution of a disk galaxy, which emphasizes the role of active galactic nucleus (AGN) feedback and incorporates cosmological inflows. This paper presents the model setup and the overall results. The predicted AGN duty cycle of $\sim 0.49\%$ is consistent with observations. Analysis of the AGN luminosity and star formation rate (SFR) light curves reveals a positive correlation between the two. We find that cold filaments condense in the circumgalactic medium (CGM) region due to radiative cooling and subsequently fall onto the galaxy, significantly enhancing both the SFR and AGN activity. The galaxy is then quenched over a timescale of $\sim 1~{\rm Gyr}$ by the strong feedback from the enhanced AGN activity. This indicates that a positive correlation between SFR and AGN luminosity does not preclude AGN feedback from acting as the quenching mechanism for the galaxy. Notably, models without AGN feedback exhibit significantly lower peak SFRs than those with it. We attribute this difference to cumulative AGN feedback, which drives gas from the galaxy into the CGM, facilitating the formation of more massive cold filaments and ultimately promoting more intense starburst episodes. 
\end{abstract}

\keywords{Galaxies: evolution — Galaxies: spiral — Galaxies: active — 
          Galaxies: nuclei — Galaxies: star formation — Galaxies: quenching — ISM: jets and outflows — Methods: numerical }

\section{Introduction}

A wealth of observations has established tight correlations between the masses of supermassive black holes (SMBHs) and the properties of their host galaxies, including stellar mass, bulge luminosity, and stellar velocity dispersion (e.g., \citealt{Magorrian1998, Ferrarese_2000, Gebhardt_2000, Kormendy_ho_2013}), suggesting coevolution between galaxies and their central SMBHs. Among the proposed physical mechanisms linking the two, active galactic nucleus (AGN) feedback is widely regarded as a leading candidate (e.g., \citealt{silk1998quasars, King_2003, Murray_2005, Fabian2012, Harrison2018}).

Despite this broad consensus, the sign and overall impact of AGN feedback on galaxy evolution—whether positive, negative, or effectively neutral—remains an active topic of debate. This controversy largely stems from seemingly contradictory observational results. On the one hand, studies have reported elevated star formation rates (SFRs) in typical AGN hosts relative to non-AGN galaxies (e.g., \citealt{Koss_2011, santini_2012, Ellison_2016, Woo_2020,Xie_2021,Molina_2023}), and evidence for positive feedback has also been identified in radio galaxies, where AGN luminosity and specific SFR (sSFR) are positively correlated (e.g., \citealt{Shen2020}). On the other hand, other works have found that SFRs in AGN hosts are lower than those in matched non-AGN control samples (e.g., \citealt{Schawinski_2007, Silverman_2008, Mullaney_2015, sanchez2018ssdss, Lacerda_2020}), consistent with negative feedback. There is also evidence that star formation in typical AGN hosts is statistically consistent with that of non-AGN galaxies at the same evolutionary stage (e.g., \citealt{Bongiorno2012, Chang_2017, Suh_2017, Nascimento_2019}), suggesting an effectively neutral (net-zero) feedback effect.

In principle, luminous AGN can release sufficient energy to unbind or expel the interstellar medium (ISM) gas from their host galaxies (e.g., \citealt{silk1998quasars, King_2003, Murray_2005}). Observations of quasar outflows indicate that their kinetic power can be high enough to drive substantial AGN feedback (e.g., \citealt{Liu2013, Shen2023}). Meanwhile, integral-field spectroscopy from MaNGA \citep{Lammers2022} suggests that AGN feedback can strongly suppress star formation in galactic centers while being comparatively inefficient at quenching star formation on global (i.e., whole-galaxy) scales. On the theoretical side, numerous hydrodynamic simulations have been carried out to investigate the role of AGN feedback, spanning both idealized, non-cosmological simulations of individual systems (ranging from galaxies to galaxy groups and clusters) \citep[e.g.,][]{2012MNRAS.424..190G,2013ApJ...768...11B,LiBryan2014,2015ApJ...811...73L,Prasad2020,Meece2017,Prasad2026} and cosmological simulations \citep{matteo2005energy, Di_Matteo_2005,Springel05,cattaneo2006modelling, croton2006the,puchwein2012shaping,Choi_2012, Hopkins_2016, Anglés-Alcázar_2021,vogelsberger2013a}. Several large and influential cosmological simulation projects include EAGLE \citep{2015MNRAS.446..521S}, IllustrisTNG \citep{Weinberger2017}, COLIBRE \citep{2025arXiv250905179H}, and SIMBA \citep{2019MNRAS.486.2827D}.

In this work, we investigate the evolution of a disk galaxy using the MACER framework \citep{2018ApJ...857..121Y}. This approach employs an idealized simulation designed to isolate and examine the role of AGN feedback in the evolution of an individual galaxy. The model builds on a series of earlier studies \citep[e.g.,][]{2001Ciotti,2009Ciotti,2011Novak,2014Gan}, and a recently updated three-dimensional version is presented in \citet{2025ApJ...985..178Z}. 
Compared with the aforementioned theoretical models, MACER has some distinctive features. A detailed comparison is presented in \citet{2025arXiv251102796H}. Here we briefly summarize the salient features.

\begin{itemize}
\item Black hole accretion rate. In MACER, the inner boundary of the simulation lies within the Bondi radius, which corresponds to the outer boundary of the accretion flow. By combining the mass flux measured at the inner boundary with black hole accretion theory, we obtain a physically motivated and comparatively robust estimate of the mass accretion rate at the event horizon, which directly determines the AGN power. In contrast, the accretion rate in most other studies is estimated in a much more approximate manner. In idealized simulations, it is commonly computed by dividing the total cold gas mass within an arbitrarily chosen radius by an assumed accretion timescale. In cosmological simulations, Bondi or modified Bondi prescriptions are typically adopted. However, since the Bondi radius is typically unresolved and the Bondi accretion rate formula can only be applied at the Bondi radius, this approach can lead to  significant overestimates or underestimates of the black hole accretion rate (e.g., \citealt{Negri2017,Hopkins_2016}).
\item Properties of AGN outputs: jets, wind, and radiation. In MACER, the properties of AGN outputs are specified as functions of the accretion rate and are either adopted from state-of-the-art black hole accretion theory or constrained by statistical observational results. For instance, in the low-accretion (hot) mode, the mass flux, velocity, and opening angle of both jets and winds are taken from 3D GRMHD simulations of black hole accretion, while in the cold mode the wind properties are adopted from observations. In contrast, idealized numerical simulations typically neglect winds altogether and treat jet properties as free parameters. As a result, key jet characteristics, such as the opening angle, are often assumed to be several times larger than those inferred from observations or GRMHD simulations. The treatment of jets and winds in cosmological simulations is highly diverse. In EAGLE and IllustrisTNG, collimated jets and accretion-driven winds are not modeled explicitly: AGN feedback is implemented via isotropic thermal energy injection in EAGLE and in the cold mode of IllustrisTNG. Although IllustrisTNG employs kinetic energy injection in the hot mode, this feedback remains phenomenological and does not explicitly model collimated jets or physically motivated accretion-driven winds. SIMBA includes both radiative-mode winds and jet-mode feedback, but these outflows are also implemented phenomenologically, with their properties calibrated to observations rather than derived from black hole accretion physics.
\item The interactions between AGN outputs and the interstellar medium (ISM). In MACER, AGN outputs are injected at the inner boundary of the simulation, and their interactions with the ISM are calculated self-consistently, including both momentum and energy exchange. In contrast, idealized simulations often treat accretion-driven winds in a highly simplified way, or neglect their coupling to the ISM altogether.
In many cosmological simulations, AGN feedback is instead implemented in a “thermal feedback” mode, in which energy is deposited isotropically into the surrounding gas within a prescribed radius from the black hole. In such models, both the fraction of AGN energy coupled to the gas and the deposition radius are typically calibrated and treated as free parameters. Moreover, the momentum feedback associated with AGN winds, which can be important in some regimes relative to pure energy feedback (e.g., \citealt{2010ApJ...722..642O}), is often neglected.
\end{itemize}

Our previous applications of MACER have focused on elliptical galaxies \citep{2018ApJ...857..121Y,Yoon_2018,2018ApJ...866...70L,2019ApJ...885...16Y,2023MNRAS.524.5787Z,2023MNRAS.525.4840Z}, compact galaxies \citep{2023MNRAS.523.1641D}, dwarf galaxies \citep{2025arXiv251020897S}, and galaxy clusters \citep{he2025}. The majority of galaxies in the universe are disk galaxies; thus, understanding the evolution of disk galaxies and how AGN feedback operates within them is essential for addressing key questions, such as the drivers of quenching and the connections to other properties (e.g., stellar mass and kinematic heating; \citealt{mo2023}).

As part of our series on the role of AGN feedback in galaxy evolution, this paper and several subsequent papers begin a systematic investigation of AGN feedback in the evolution of a disk galaxy within the MACER framework. The present paper describes the model setup and presents the general results, including the long-term AGN light curve, the predicted AGN duty cycle and its comparison with observations, the history of the star formation rate, and a comparison between models with and without AGN feedback. Subsequent papers will address: 1) the correlation between the black hole accretion rate and the star formation rate; 2) the radial profile of the X-ray surface brightness predicted by the model and its comparison with \textit{eROSITA} observations; 3) the quenching of the galaxy and its physical mechanisms; 4) the (cold) gas fraction; and 5) the correlation between black hole mass and velocity dispersion.

This paper is organized as follows. In Section~\ref{sec:Model}, we describe the numerical setup and the physical models used in our simulations. The main results are presented in Section~\ref{sec:Results}. We discuss and summarize our findings in Section~\ref{sec:Discussion}.

\section{Model}
\label{sec:Model}

The details of the MACER framework can be found in \citet{2018ApJ...857..121Y} (hereafter \citetalias{2018ApJ...857..121Y}). In this paper, we briefly summarize the most salient features of the MACER model and describe several new components introduced in this work, including the simulation initial conditions, our new treatment of supernova (SN) explosions, the cosmological gas inflow, and the AGN jet prescription.
  
\subsection{Hydrodynamics}

Our galaxy evolution, which includes physical processes such as star formation, AGN feedback, and stellar feedback, is computed by integrating the following time-dependent Navier–Stokes equations written in Eulerian form for the conservation of mass, momentum, and thermal energy:
\begin{equation}
    \frac{\partial \rho}{\partial t} + \nabla \cdot (\rho\ v) = \alpha_\mathrm{*}\,\rho_\mathrm{*} + \dot{\rho}_\mathrm{SN} - \dot{\rho}^{+}_\mathrm{SF},
\end{equation}
\begin{equation}
    \frac{\partial \bm{m}}{\partial t} + \nabla \cdot (\bm{m}\,\bm{v}) = -\nabla p_\mathrm{gas} + \rho\,\bm{g} + \dot{\bm{m}}_\mathrm{SN} - \dot{\bm{m}}^{+}_\mathrm{SF} + \nabla \cdot \bm{T},
\end{equation}
\begin{equation}
    \frac{\partial E}{\partial t} + \nabla \cdot (E\,\bm{v}) = -p_\mathrm{gas}\,\nabla \cdot \bm{v} + H - C + \dot{E}_\mathrm{S} + \dot{E}_\mathrm{SN} - \dot{E}^{+}_\mathrm{SF} + \bm{T}^2/\mu.
\end{equation}
Here, \(\rho\), \(\bm{m}\), and \(E\) denote the local gas mass density, momentum density, and thermal energy density, respectively; the gas pressure is given by \(p_\mathrm{gas} = (\gamma - 1)E\). The terms \(\alpha_\mathrm{*}\rho_\mathrm{*}\) and \(\dot{E}_\mathrm{S}\) represent the mass and thermal energy injection by stellar winds; the subscript “SN” denotes mass, momentum, and energy input from supernovae, while the superscript “+” denotes sinks due to star formation. \(H\) and \(C\) are the radiative heating and cooling rates arising from interactions between radiation and the ISM (see \citetalias{2018ApJ...857..121Y} for details). \(\bm{g}\) is the gravitational acceleration. The tensor \(\bm{T}\) is the viscous stress tensor, used here to approximate the angular momentum transfer, and \(\mu\) is the dynamic viscosity. The final term represents viscous dissipation.We will describe the physics related to the stress tensor $T$ in detail in \S\ref{sec:angularmomentum}.

Following \citetalias{2018ApJ...857..121Y}, we perform two-dimensional, axisymmetric simulations with the parallel ZEUS-MP/2 code \citep{Hayes_2006} in spherical polar coordinates \((r,\theta,\phi)\). The angular mesh consists of 72 uniformly spaced zones in \(\theta\), while the radial mesh spans from \(100~\mathrm{pc}\) to \(500~\mathrm{kpc}\) with 280 logarithmically spaced zones. To avoid the coordinate singularity on the polar axis, we excise a narrow cone around \(\theta=0\). With this discretization, the smallest cell size occurs at the inner radial boundary and is approximately \(4~\mathrm{pc}\).

\subsection{Initial conditions of the disk galaxy}
We consider the evolution of a disk galaxy with the inclusion of cosmological inflows. The galaxy is modeled with several components: a central supermassive black hole (SMBH),  a rotationally supported central stellar bulge, both stellar and gaseous disks, and a dark matter halo. These components are embedded within the overall gravitational structure of the galaxy. The self-gravity of the interstellar medium (ISM) is neglected. The background gravitational potential that governs the dynamics is:
\begin{equation}
    \Phi = \phi_{\mathrm{BH}} + \phi_{\mathrm{disc,\ast}} + \phi_{\mathrm{bulge}} + \phi_{\mathrm{DM}}.
\end{equation}

The circular-velocity and density profiles of the dark matter halo, stellar bulge, and stellar disk follow those in \citet{Springel05}. The initial gas disk is described by the softened profile of \citet{Tonnesen_2009}:
\begin{equation}
    \rho(R,z) = \frac{M_\mathrm{g}}{8\pi R_\mathrm{g}^{2} z_\mathrm{g}}\,
    \mathrm{sech}\!\left(\frac{R}{R_\mathrm{g}}\right)\,
    \mathrm{sech}\!\left(\frac{z}{z_\mathrm{g}}\right),
\end{equation}
within $R \le R_{\max}$ and $|z| \le z_{\max}$, where $R_{\max} = 24\,\mathrm{kpc}$ and $z_{\max} = 2\,\mathrm{kpc}$ denote the edges of the gaseous disk, which also mark the interface between the ISM and the circumgalactic medium (CGM). We set the initial gas disk scale length $R_\mathrm{g} = 3.5\,\mathrm{kpc}$ and scale height $z_\mathrm{g} = 0.325\,\mathrm{kpc}$. The initial CGM mass is $2.0\times10^{10}\,M_{\odot}$. 

The initial azimuthal velocity of the gas disk is taken to be rotationally supported:
\begin{equation}
    v_{\phi,\mathrm{gas}}^{2} = R\,\frac{\partial \Phi}{\partial R},
\end{equation}
neglecting pressure support. We adopt an ideal-gas equation of state with an adiabatic index
\begin{equation}
    \gamma \equiv \frac{d \ln P}{d \ln \rho} = \frac{5}{3}.
\end{equation}

\begin{figure*}
\centering
\includegraphics[width=\textwidth]{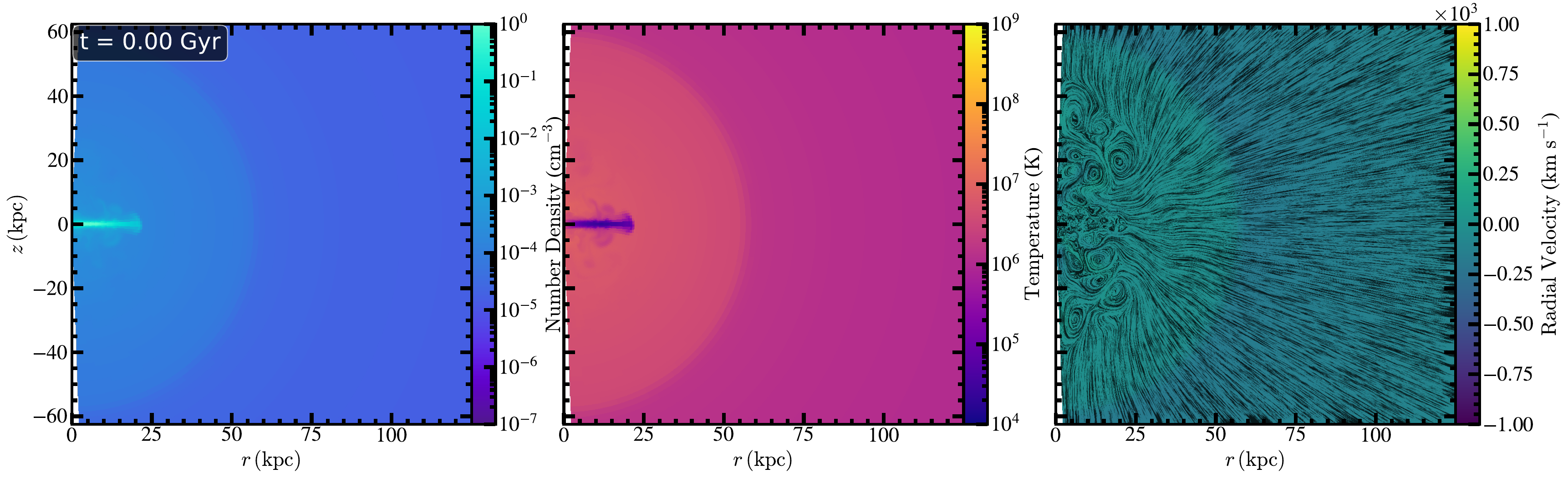}
\caption{Initial conditions for the simulations: spatial distributions of the gas number density, temperature, and radial velocity. These fields are taken from the end of a 0.5625~Gyr relaxation run, during which the system evolves from the imposed initial setup toward an approximately hydrostatic configuration. We define this snapshot as \(t = 0\) and adopt the resulting state as the starting point for all subsequent simulations.}
\label{fig:init}
\end{figure*}

Our disk galaxy is modeled as a system with halo parameters \((M_{\mathrm{halo}}, c) = (1.6 \times 10^{12}\, M_{\odot},\, 12)\), where \(M_{\mathrm{halo}}\) is the halo mass and \(c\) is the concentration parameter. The masses of the galactic components are \((M_{b}, M_{\ast}, M_{g}) = (1.5 \times 10^{10}\, M_{\odot},\, 4.7 \times 10^{10}\, M_{\odot},\, 0.9 \times 10^{10}\, M_{\odot})\) for the bulge, stellar disk, and gas disk, respectively. The stellar disk has scale lengths \((h_{\ast}, z_{\ast}) = (3.0\, \mathrm{kpc},\, 0.3\, \mathrm{kpc})\), following the ``Milky Way'' model of \citet{Hopkins2012}. The central supermassive black hole has an initial mass of \(5 \times 10^{7}\, M_{\odot}\). The initial ISM temperature is set to \(10^{4}\, \mathrm{K}\), while the CGM is initialized at \(10^{6}\, \mathrm{K}\). We allow the initial conditions to relax self-consistently for \(0.5625\, \mathrm{Gyr}\) to establish hydrostatic equilibrium, after which the production run begins (Fig. \ref{fig:init}).
Our simulation spans 12\,Gyr, implying a starting epoch corresponding to \(z \gtrsim 3\). 

We emphasize that the gas density adopted here is calibrated to low-redshift disk galaxies and is therefore lower than the typical values expected at such high redshifts. As a result, during the early stages of our simulated evolution, the SFR is noticeably lower than the observational values reported at \(z \sim 2-3\).  We do not adopt a more realistic high-\(z\) density because our simulations remain somewhat idealized: they do not account for several important physical processes, such as galaxy mergers and the evolution of the galactic gravitational potential driven by the cosmological growth of dark matter halos. Our goal is to isolate and investigate the key physical mechanisms that regulate galaxy evolution, rather than to achieve a precise quantitative match to observations. This limitation should therefore be kept in mind when comparing the absolute values of our predicted quantities with observational data.

\subsection{Cosmological inflows}

We model the environmental gas supply as a combination of hot- and cold-mode cosmological inflows. Parameter choices follow \citet{Dekel09}. The total inflow rate follows the baryonic growth-rate scaling of \citet{Dekel09}:
\begin{equation}
\dot{M} \approx 6.6\,M_{12}^{1.5}\,(1+z)^{2.25}\,f_{0.165}\; M_\odot\,\mathrm{yr}^{-1},
\end{equation}
where \(M_{12}\equiv M_{\rm vir}/10^{12}M_\odot\) and \(f_{0.165}\equiv f_b/0.165\). For \(M_{\rm vir}=1.6\times 10^{12}\,M_\odot\), adopting a representative \(z\simeq 1.5\) and \(f_{0.165}\simeq 1\) yields \(\dot{M}\simeq 1.05\times 10^{2}\,M_\odot\,\mathrm{yr}^{-1}\), which we round to \(100\,M_\odot\,\mathrm{yr}^{-1}\) and impose as a fixed boundary condition despite the expected cosmological evolution of accretion rates.

We partition the inflow in a 60:40 ratio between the cold and hot modes, motivated by the cold-mode dominance in lower-mass halos that declines with mass and approaches parity near \(M_{\rm halo}\sim 10^{11.4}\,M_\odot\) \citep{Keres05}, and by the persistence of cold streams as the primary gas supply for \(M_{\rm vir}\sim 10^{12}\,M_\odot\) halos at high redshift \citep{Dekel09}. Given our target halo mass \(\left(1.6\times 10^{12}\,M_\odot\right)\) in this transitional regime, we  adopt cold-mode dominance (60\%) with a substantial hot component (40\%).

The two modes are implemented with distinct approaches. A quasi-spherical hot-mode inflow is injected at \(R=500\,\mathrm{kpc}\) with \(T_{\rm hot}=1.0\times 10^6\,\mathrm{K}\) and  \(n_{\rm hot}=1.0\times 10^{-5}\,\mathrm{cm^{-3}}\). The cold-mode filamentary inflows are injected  at \(R=100\,\mathrm{kpc}\) with \(T_{\rm cold}=1.1\times 10^4\,\mathrm{K}\) and \(n_{\rm cold}=2.9\times 10^{-4}\,\mathrm{cm^{-3}}\). The choice of the injection radius of the cold-mode inflow is physically motivated by the survival of cold streams in this region and their ability to feed the inner disk (\(\sim 15\,\mathrm{kpc}\); \citealt{Dekel09}). In our simulations, directly imposing narrow cold filaments at the outer boundary of the domain ($r=500$ kpc) would cause them to become numerically diffused and dissolved before reaching the inner halo, owing to our limited resolution at large radii. 

\citet{Dekel09} showed that cold accretion in halos of mass \(M_{\rm vir}\sim 10^{12}\,M_\odot\) at high redshift is dominated by a small number (\(\sim 3\)) of narrow, radially oriented filaments that penetrate the hot halo and are distributed over a wide range of inclinations relative to the galactic disk.
Following this picture, the cold-mode inflow in our model is implemented in the form of three localized, filamentary streams with finite opening angles rather than as a spherical component. The three filaments in our simulations are oriented approximately radially and are inclined with respect to the disk midplane, with characteristic polar angles of order \(\sim 20^\circ\) above the midplane, \(\sim 25^\circ\) below the midplane, and \(\sim 70^\circ\) below the midplane. Each filament subtends a finite angular width of order \(\sim 8^\circ\!-\!12^\circ\), corresponding to physical widths of \(\sim 16\!-\!20\,\mathrm{kpc}\) at the injection radius. In an axisymmetric geometry, such three-dimensional cold streams appear as extended arc-like inflow regions in the \((r,\theta)\) plane.
A constant radial inflow velocity of \(v_r=-200\,\mathrm{km\,s^{-1}}\) (negative denotes inflow) is imposed on these filaments, consistent with characteristic cold-stream velocities in massive halos at high redshift. In contrast, the hot-mode inflow is not assigned an explicit radial velocity. Instead, gas entering through the spherical boundary at \(R=500\,\mathrm{kpc}\)  inherits the existing local radial velocity field at that radius. This approach thereby allows hot-mode accretion to self-consistently respond to the global dynamics of the halo.
The specific angular momentum of the cold-mode inflow is set via the azimuthal velocity \(v_\phi\), with two limiting cases considered: a high-angular-momentum inflow \(\left(v_\phi=200\,\mathrm{km\,s^{-1}}\right)\) and a low-angular-momentum inflow \(\left(v_\phi\approx 0\right)\). No azimuthal velocity is imposed on the hot-mode inflow. We note that our approach to modeling the cold-mode inflow has certain limitations, as it inevitably influences the hydrodynamics of the AGN outflow. This interaction should be considered when interpreting the results.

\subsection{Angular momentum transfer}
\label{sec:angularmomentum}

Angular momentum transfer in galaxies is an important but challenging problem. Many mechanisms have been proposed over the past several decades, including tidal torques during gas-rich mergers \citep{1989Natur.340..687H}, magnetorotational instability in galactic disks \citep{1999ApJ...511..660S}, turbulence driven by star-formation feedback and associated radial transport \citep{2018MNRAS.477.2716K}, instabilities in self-gravitating disks (bars and spiral density waves) \citep{2008ApJ...686..815Y}, and global gravitational torques (``bars within bars'') \citep{1989Natur.338...45S,Hopkins2010,Hopkins2011}.

Since our simulation is two-dimensional, it is difficult to self-consistently consider the angular-momentum transport processes in our work. On the other hand, the exact transfer mechanism may not be the sole determinant of key quantities such as the black hole accretion rate and the star formation rate. Once the inflow rate driven by the angular momentum transfer mechanism is sufficiently high, the black hole accretion rate and the gas surface density are largely set by the interaction between the AGN and the ISM; in other words, these quantities are self-regulated \citep[e.g.,][]{Weinberger2017}. This self-regulation diminishes the sensitivity to the details of angular-momentum transport. Nevertheless, as we demonstrate in this paper, different transport efficiencies in our models do lead to quantitatively different outcomes.

Given these reasons, we formally introduce an anomalous stress tensor to approximate the effects of physical angular momentum transport\footnote{A similar approach was adopted in the canonical analytical ``$\alpha$-disk'' work \citep{1973A&A....24..337S} and in the numerical simulation study \citep{1999MNRAS.310.1002S} of black hole accretion disks, when the magnetic field—responsible for angular momentum transport—is not explicitly included in the models.}. For spherical polar coordinates \((r,\theta,\phi)\), given that the gas rotation is primarily in the \(\phi\)-direction, the relevant components of the viscous stress tensor we adopt are
\begin{equation}
T_{r\phi} \;=\; \mu\, r\, \frac{\partial}{\partial r}\!\left(\frac{v_\phi}{r}\right), 
\qquad
T_{\theta\phi} \;=\; \mu\, \frac{\sin\theta}{r}\, \frac{\partial}{\partial \theta}\!\left(\frac{v_\phi}{\sin\theta}\right),
\end{equation}
where \(\mu \equiv \nu \rho\) is the dynamic viscosity and \(\nu\) is the kinematic viscosity. The form and magnitude of \(\nu\) set the efficiency of angular momentum transport. In this work, for simplicity, we adopt a spatially uniform kinematic viscosity, \(\nu=\mathrm{const}\). By varying the value of \(\nu\) across models, we aim to roughly mimic different mechanisms or different magnitudes of angular momentum transfer that may operate in different disk galaxies. 
In this context, higher kinematic viscosity effectively models systems in which angular momentum transport is dominated by efficient, small-scale processes, such as turbulence driven by star-formation feedback, magnetorotational instability, or unresolved gravitational torques, leading to relatively smooth and rapid radial redistribution of angular momentum. In contrast, lower kinematic viscosity represents systems where angular momentum transport is weaker or more intermittent and is governed primarily by large-scale, coherent dynamical processes, such as bars, spiral density waves, or tidal torques during mergers, rather than by local diffusive transport.
In addition, varying the values of $\nu$ serves another purpose. In this series of work, we use different snapshots of a single simulated galaxy to represent a population of disk galaxies, an approach that inevitably has limitations. Exploring different values of $\nu$ helps enhance the diversity of the resulting galaxy properties. For these reasons, four values of $\nu$ are considered in the present study, namely \(\nu = 0.25, 0.55, 0.75,\) and \(1.1\). 

\begin{figure}
\centering
\includegraphics[width=0.9\columnwidth]{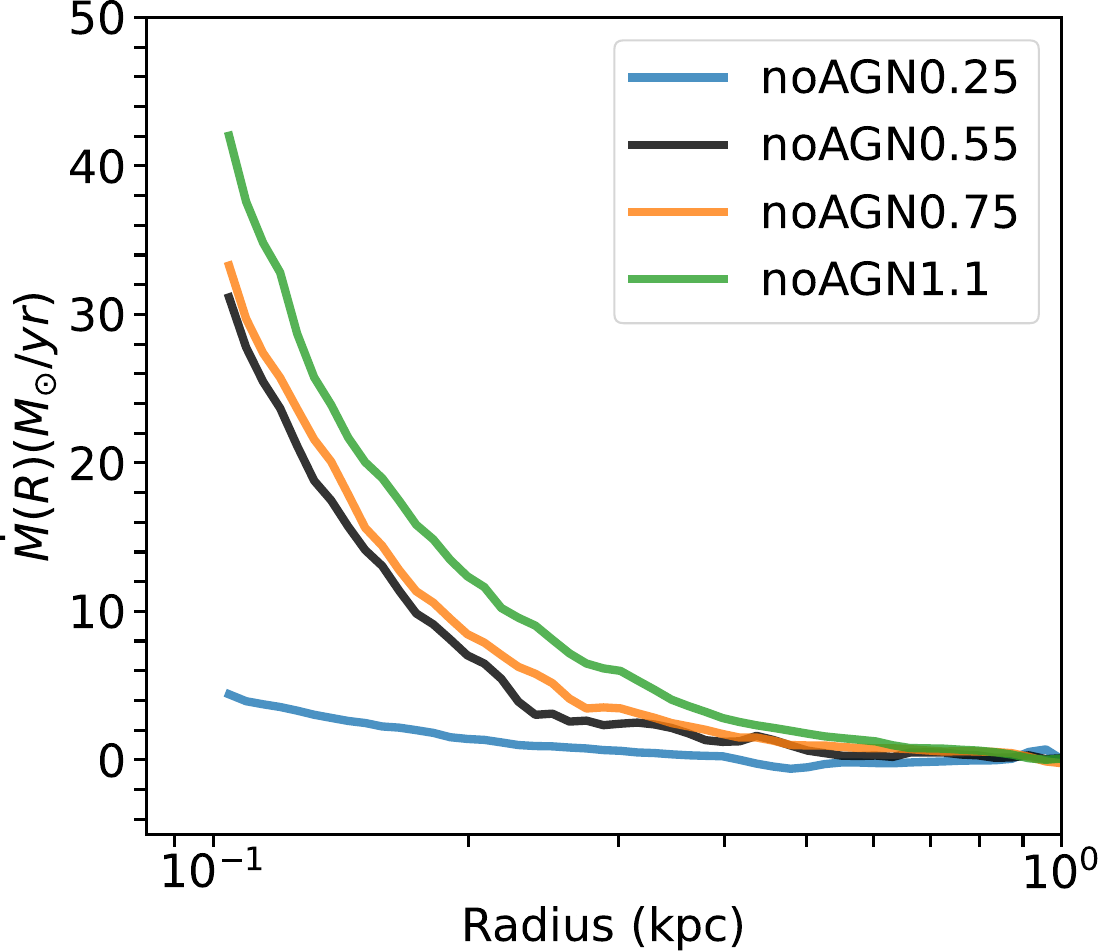}
\caption{Time-averaged mass inflow  rate in the inner region of the galaxy as a function of radius for various values of viscous parameter $\nu$. AGN feedback is turned off. }
\label{fig:gas_inflow}
\end{figure}

Figure \ref{fig:gas_inflow} shows the radial profiles of the mass inflow rate for different choices of \(\nu\) in our models; AGN feedback is not included in these calculations. We find that the inflow rates are roughly proportional to the magnitude of \(\nu\). Since the radial velocity scales with \(\nu\), this implies that the surface density profiles of the various models are roughly the same, and thus the star formation rates should also be similar. However, we caution that varying the parameter values in this manner can only partially probe the true angular momentum transport, because the radial dependence of the viscosity is held fixed, and it is uncertain whether this scaling represents real systems.

We now compare the resulting inflow rates in our models to the numerical simulations of angular momentum transport by \citet{Hopkins2010,Hopkins2011}. In these two studies, the inflow of gas is followed from galactic scales (\(\sim 10\,\mathrm{kpc}\)) down to \(\lesssim 0.1\,\mathrm{pc}\) using hydrodynamical simulations. For sufficiently gas-rich, disk-dominated systems---similar to our present case---they find that a sequence of gravitational instabilities generates large accretion rates, with the strongest torques on the gas arising when non-axisymmetric perturbations to the stellar gravitational potential produce orbit crossings and shocks in the gas. The inflow rates obtained in their simulations are highly variable in both time and space. At \(\sim 100\,\mathrm{pc}\) (which is the inner boundary of our simulation domain), the typical inflow rate is of order tens of \(M_\odot\,\mathrm{yr}^{-1}\). We note that the inflow rates at \(100\,\mathrm{pc}\) predicted by the ``noAGN0.55'', ``noAGN0.75'', and ``noAGN1.1'' models are similar to this value, as shown in Figure \ref{fig:gas_inflow}.

A useful observational criterion for assessing whether the angular momentum transfer adopted in the model is reasonable is to compare theoretical predictions of key quantities, such as the black hole accretion rate (BHAR) and the star formation rate (SFR), with observations. As we show in this paper and in subsequent papers, both BHAR and SFR are in good agreement with observations. In particular, as we will describe later, our fiducial model yields an AGN duty cycle (defined as the ratio of the duration spent in the cold mode to the total AGN duration) that is well consistent with observational constraints\footnote{This is why we refer to that model as ``Fiducial''.}. Nevertheless, in future work we plan to adopt different prescriptions for angular momentum transfer and investigate their effects.

\subsection{AGN feedback}
\label{subsec:agn-feedback}

The inner boundary of our simulation domain is smaller than the Bondi radius, which marks the outer boundary of the black hole accretion flow. Within the Bondi radius, accretion onto the black hole is treated with subgrid physics. Following \citetalias{2018ApJ...857..121Y}, we implement a two-mode AGN feedback model comprising a cold (quasar) mode at high accretion rates and a hot (radio/kinetic) mode at low accretion rates. The boundary between the two modes is set by a critical luminosity \(L_{\rm c} = 0.02\,L_{\rm Edd}\) \citep{McClintock2006,Yuan2014}. The corresponding critical accretion rate is
\begin{equation}
\dot{M}_{\rm c} \equiv \frac{L_{\rm c}}{\epsilon_{\rm EM,cold}\,c^{2}},
\end{equation}
where \(\epsilon_{\rm EM,cold}\) is the radiative efficiency of a geometrically thin, radiatively efficient accretion disk. In our implementation, we adopt \(\epsilon_{\rm EM,cold} = 0.1\). We measure the mass inflow rate at the inner boundary  and then compare \(\dot{M}(r_{\rm in})\) against \(\dot{M}_{\rm c}\) to determine the AGN mode.

\subsubsection{The cold accretion flow regime}
The cold mode is further divided into the standard thin disk and super-Eddington accretion, when the mass accretion rate is lower or higher than the Eddington rate, respectively. The dynamics of the accreting gas within the Bondi radius is complex and warrants future study. In MACER, we adopt the following simplified scenario: the gas first free-falls until an accretion disk forms at the circularization radius. We consider  wind launched from the disk. The effective accretion rate at which gas feeds the accretion disk is obtained by solving
\begin{equation}
\frac{d\dot{M}_{\rm eff}}{dt}
= \frac{\dot{M}(r_{\rm in}) - \dot{M}_{\rm eff}}{\tau_{\rm ff}},
\label{eq:feed_lag}
\end{equation}
where \(\tau_{\rm ff}\) is the free-fall timescale, 
$\dot{M}(r_{\rm in})$ is the mass inflow rate at the inner boundary $r_{\rm in}$, $\tau_{\rm ff}$ is the free-fall time at $r_{\rm in}$. Gas fuels the black hole on a viscous timescale \(\tau_\mathrm{vis}\) (\citetalias{2018ApJ...857..121Y}, Eqs.~6). With the computed total mass of the gas in the accretion disk $M_{\rm dg}$, the mass inflow rate at $R_{\rm cir}$ is estimated as 
\begin{equation}
\dot M_{d,\mathrm{inflow}} = \frac{M_{dg}}{\tau_\mathrm{vis}}.
\label{eq:viscous}
\end{equation}
The black hole accretion rate is equal to the disk inflow rate reduced by the mass carried away by the  wind,
\begin{equation}
\dot M_\mathrm{BH} = \dot M_{d,\mathrm{inflow}} - \dot M_{\mathrm{w,cold}}. 
\end{equation}
The disk mass evolves as
\begin{equation}
\frac{\mathrm{d}M_{dg}}{\mathrm{d}t} = \dot {M}^\mathrm{eff} - \dot M_\mathrm{BH} - \dot M_{\mathrm{w,cold}}.
\label{eq:bh_and_diskmass}
\end{equation}
 Combining Equations (\ref{eq:feed_lag}) – (\ref{eq:bh_and_diskmass}), we can estimate the black hole accretion rate $\dot{M}_{\rm BH}$.  The mass flux and velocity of cold-disk wind are prescribed based on observations as a function of the AGN bolometric luminosity (\citetalias{2018ApJ...857..121Y}, Eqs.~9--12; \citealt{Gofford2015}). The wind mass flux is distributed in angle as \(\propto \cos^{2}\theta\). The radiation luminosity $L_{\rm bol}=\epsilon_{\rm EM,cold}\dot{M}c^2$. 
 
 The super-Eddington regime is neglected in \citet{2018ApJ...857..121Y} but is included here (see also \citep{2023MNRAS.524.5787Z}). We consider wind and radiation but neglect the possible jet in this regime. The properties of the mass flux and velocity of wind are taken from \citet{2023MNRAS.523..208Y} in which they analyzed the three-dimensional general relativity radiative MHD simulations  of super-Eddington accretion around a supermassive black hole. The radiation property is taken from the simulations of super-Eddington accretion flow presented in \citet{2019ApJ...885..144J}.  

\subsubsection{The hot accretion flow regime}

In the hot mode, we consider radiation, winds, and jets. The accretion proceeds via a hot accretion flow inside a truncation radius \(r_\mathrm{tr}\), exterior to which a thin disk exists \citep{Yuan2014}. 
Taking \(\dot M(r_\mathrm{in})\approx \dot M(r_\mathrm{tr})\) and following the theory of winds from hot accretion flows \citep{Yuan2015}, the black hole accretion rate is
\begin{equation}
\dot M_{\mathrm{BH,hot}}\;\approx\;\dot M(r_\mathrm{tr})\left(\frac{3r_s}{r_\mathrm{tr}}\right)^{1/2}.
\label{eq:mdot_bh_hot}
\end{equation}
The wind mass flux and velocity are prescribed by \citep{Yuan2015}
\begin{equation}
    \dot{M}_\mathrm{w,hot} \;\simeq\; \dot{M}(r_\mathrm{in})\left[\,1- \left(\frac{3\,r_\mathrm{s}}{r_\mathrm{tr}}\right)^{1/2}\right],
    \label{eq:mdot_w_hot}
\end{equation}
\begin{equation}
    v_\mathrm{w,hot} \;=\; 0.2\,v_\mathrm{K}(r_\mathrm{tr}),
    \label{eq:vw_hot}
\end{equation}
where \(v_\mathrm{K}(r_\mathrm{tr})\) is the Keplerian speed at \(r_\mathrm{tr}\). The angular distribution of the hot-mode wind is confined to intermediate polar angles, following \citetalias{2018ApJ...857..121Y}: \(\theta\sim 10^{\circ}\)-- \(40^{\circ}\) and \(140^{\circ}\)-- \(170^{\circ}\). Radiative output in the hot mode is computed with a radiative efficiency for hot accretion flows, \(\epsilon_\mathrm{EM,hot}=\epsilon_\mathrm{hot}(\dot{M}_\mathrm{BH})\), following the fitting formulae in \citet{Xie2012}.


We now introduce the implementation of jet in the hot mode. Based on calculations of the energy and spatial distribution of energetic electrons accelerated by magnetic reconnection in the jet and their resulting radiation, \citet{2024SciA...10N3544Y} show that the observed AGN jet should be described by the Blandford-Znajek model \citep{1977MNRAS.179..433B}, which has received intensive studies over the last decade \citep{Yuan2014}. However, in almost all existing literature of AGN feedback, the jet parameters are typically treated as unconstrained free variables, whose values are not necessarily consistent with the constraints obtained from GRMHD simulations of Blandford-Znajek jet. In the Blandford-Znajek model, the properties of jets are  determined by black hole spin and the accretion mode (SANE or MAD).  We assume a high dimensionless spin parameter and MAD in the present work. Under these assumptions, the jet parameters are taken from the three-dimensional general relativity MHD simulations of  black hole accretion of \cite{Yanghai_2021}. Since the outer boundary of \citet{Yanghai_2021} is much smaller than the inner boundary of our simulation domain, extrapolation of  simulation results is required for implementation in MACER. Specifically, in our model, the jet is injected at the inner boundary of simulation domain with a speed of \(v_j = 0.5\,c\). The half-opening angle of jet is defined as \(\theta_j \equiv \arctan(R/Z)\), where \(R\) is the jet’s cylindrical radius and \(Z\) is the axial distance from the black hole. Following the suggestions of observations, we set the half-opening angle to \(\theta_j = 2.5^\circ\). The exact value may not be crucial since the jet is perpendicular to the galactic disk. The jet mass flux is set to be  \(\dot{M}_j = 0.35\,\dot{M}_{\rm BH}\).

\subsection{Star formation and supernova feedback}

Stars are formed following the same criterion as in our previous elliptical-galaxy work \citepalias{2018ApJ...857..121Y}. But we require that stars form only when the local temperature is below $4\times10^{4}\,\mathrm{K}$ and the number density exceeds $1.0~\mathrm{cm^{-3}}$.

Previously, our treatment of supernova (SN) feedback was purely thermal. To better capture realistic galactic conditions, we now also include momentum feedback from supernovae (SNe). Specifically, we assume that one SN occurs per \(70\,M_\odot\) of newly formed stars. The location of each SN is determined probabilistically, with a weight proportional to the locally formed stellar mass:
\begin{equation}
P_i \;=\; \frac{M_{\star,i}}{\sum_j M_{\star,j}}.
\end{equation}
For every cell, once the SN trigger criterion is met (i.e., an additional \(70\,M_\odot\) of stars have formed), we draw a random number \(u \sim \mathcal{U}[0,1]\); if \(u < P_i\), the SN is realized in that cell. This procedure ensures that, on average, one SN occurs per \(70\,M_\odot\) of stars formed.

To partition the injected energy and momentum, we adopt the subgrid model of \citet{Martizzi2015}. Each SN ejects \(3\,M_\odot\) of gas into the interstellar medium (ISM) and deposits a total energy of \(10^{51}\,\mathrm{erg}\), split into thermal and kinetic components that are injected radially:
\begin{equation}
E_{\mathrm{th},0} \;=\; 6.9 \times 10^{49}\,\mathrm{erg}, 
\qquad
E_{\mathrm{kin},0} \;=\; 9.31 \times 10^{50}\,\mathrm{erg}.
\end{equation}
Following \citet{Martizzi2015}, we further assume that the effective propagation scale over which an SN can drive turbulence decreases with increasing ambient density. Accordingly, we cap the propagation (coupling) radius at \(100\,\mathrm{pc}\) for \(n \gtrsim 1\,\mathrm{cm}^{-3}\), at \(10\,\mathrm{pc}\) for \(n \gtrsim 100\,\mathrm{cm}^{-3}\), and at \(1\,\mathrm{pc}\) for \(n \gtrsim 1000\,\mathrm{cm}^{-3}\). The smallest cap is also imposed where necessary for numerical stability.

\subsection{Tracer implementation}
\label{sec:tracer}

To diagnose the origin and mixing of gas from different feedback and inflow channels, we introduce a set of passive scalar tracers in the simulation. These tracers are used solely for diagnostic purposes and do not affect the gas dynamics or thermodynamics.

Each tracer, denoted by $X_n$, is defined as a mass fraction and represents gas originating from a specific source, including AGN winds, AGN jets, stellar winds, supernova (SN) ejecta, cold-mode cosmological inflow, and hot-mode cosmological inflow. The tracers are advected with the hydrodynamical flow using the same numerical scheme as the gas density. By construction, the sum of all tracer fields satisfies
\begin{equation}
\sum_n X_n = 1
\end{equation}
in each computational cell at all times.

During mass injection events, tracer fields are updated in a strictly mass-conserving manner. For a computational cell containing a pre-existing gas mass $M$, an injected mass
\begin{equation}
\Delta M = \dot{M}\,\Delta t
\end{equation}
modifies the tracer abundances according to:
\begin{equation}
X_n^{\mathrm{new}} = \frac{X_n^{\mathrm{old}}\, M + \Delta M_n}{M + \Delta M},
\end{equation}
where $\Delta M_n$ denotes the portion of the injected mass associated with tracer $n$.

For feedback or inflow channels without a direct source term during a given event, $\Delta M_n = 0$ and the corresponding tracers are diluted by the newly injected material. For channels such as supernova feedback, the injected mass is assigned entirely to the corresponding tracer ($\Delta M_n = \Delta M$), ensuring that the newly added gas is properly tagged at the moment of injection. The tracer update is performed immediately after mass injection at each time step.

As passive scalars, the tracers are subject to the same numerical diffusion as the hydrodynamic variables, but this does not affect the global mass budget and does not compromise the identification of gas origin in terms of integrated quantities. The normalization and positivity of all tracer fields are preserved by construction.

\section{Results}
\label{sec:Results}

\begin{table}[t]
  \centering
  \setlength{\tabcolsep}{6pt}
  \renewcommand{\arraystretch}{1.2}
  \caption{
  Simulation settings adopted in this work. Models are grouped by the inclusion of AGN feedback (on/off) and by variations in the viscosity coefficient $\nu$ to explore angular-momentum transport efficiencies. The column ``Angular momentum'' indicates the specific angular momentum of the cold-mode cosmological inflows (``high'' or ``low''). This design enables a direct comparison to isolate the effects of AGN feedback on galaxy quenching and to test the strangulation quenching scenario by varying the angular momentum of the filaments in cosmological inflows.
  }
  \label{tab:sim-settings}

  \begin{tabular}{@{} l c c c @{}}
    \toprule
    Name & \(\nu\) & AGN feedback & Angular momentum \\
    \midrule
    \multicolumn{4}{c}{AGN feedback on} \\
    \midrule
    Model0.25  & 0.25 & on  & high \\
    Fiducial   & 0.55 & on  & high \\
    ModelLow   & 0.55 & on  & low  \\
    Model0.75  & 0.75 & on  & high \\
    Model1.1   & 1.10 & on  & high \\
    \midrule
    \multicolumn{4}{c}{AGN feedback off} \\
    \midrule
    noAGN0.25  & 0.25 & off & high \\
    noAGN0.55  & 0.55 & off & high \\
    noAGNLow   & 0.55 & off & low  \\
    noAGN0.75  & 0.75 & off & high \\
    noAGN1.1   & 1.10 & off & high \\
    \bottomrule
  \end{tabular}
\end{table}

\begin{table}[t]
  \centering
  \setlength{\tabcolsep}{6pt}
  \renewcommand{\arraystretch}{1.2}

  \caption{Parameter survey for AGN activity. The table reports the AGN duty cycle, time-averaged AGN luminosity, and star formation rate (SFR). The AGN duty cycle predicted by the Fiducial run is consistent with observational constraints, whereas ModelLow exceeds the observed duty cycle level by about an order of magnitude, highlighting the critical role of cold cosmological inflow in regulating the AGN duty cycle.}
  \label{tab:agn-activity}

  \begin{tabular}{@{} l c c c @{}}
    \toprule
    Name & Duty cycle (\%) & \(L/L_{\rm Edd}\) & SFR (\(M_\odot\,{\rm yr}^{-1}\)) \\
    \midrule
    Model0.25  & 0.02 & \(1.44 \times 10^{-4}\) & 0.85 \\
    Fiducial   & 0.49 & \(3.23 \times 10^{-4}\) & 3.29 \\
    ModelLow   & 7.10 & \(1.41 \times 10^{-2}\) & 2.81 \\
    Model0.75  & 0.03 & \(1.26 \times 10^{-4}\) & 6.22 \\
    Model1.1   & 0.04 & \(1.12 \times 10^{-4}\) & 0.78 \\
    \bottomrule
  \end{tabular}
\end{table}

\begin{figure*}
\centering
\includegraphics[width=0.9\textwidth]{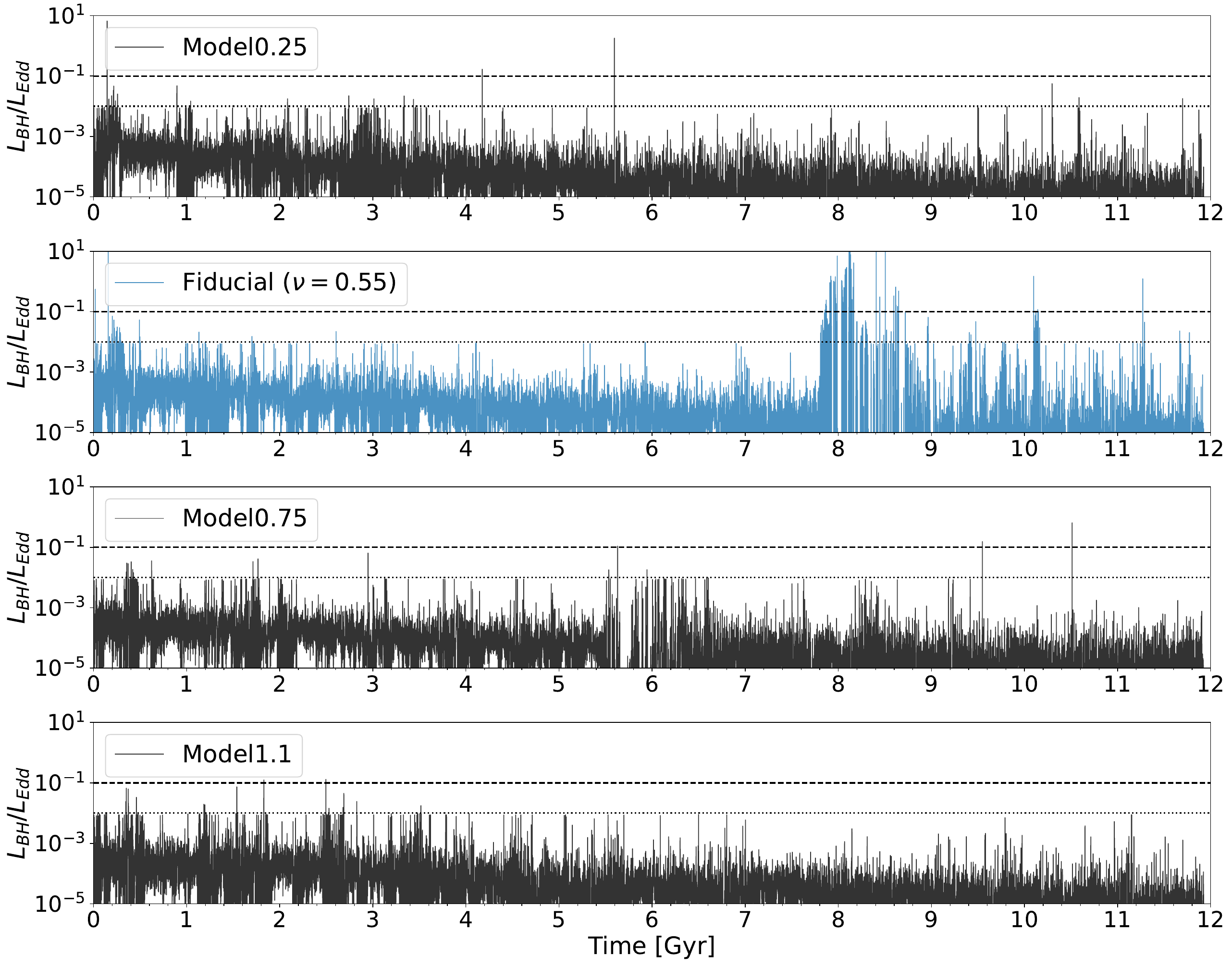}
\caption{
Time evolution of the AGN luminosity, normalized to the Eddington luminosity ($L/L_{\mathrm{Edd}}$). 
From top to bottom, the panels correspond to $\nu = 0.25, 0.55$, $0.75$, and $1.1$. 
The dotted and dashed horizontal lines denote $0.01\,L_{\mathrm{Edd}}$ and $0.1\,L_{\mathrm{Edd}}$, respectively. 
In the Fiducial model, a notable increase in AGN luminosity occurs at $t \approx 7.8~\mathrm{Gyr}$. 
An inspection of the simulation movie indicates that this rise is a direct consequence of substantial cold filaments that formed earlier in the galaxy’s CGM. 
These filaments subsequently traverse the ISM and are ultimately accreted onto the black hole, triggering AGN activity and enhancing the luminosity.
}
\label{fig:AGNlightcurve}
\end{figure*}

\begin{figure*}
\centering
\includegraphics[width=0.9\textwidth]{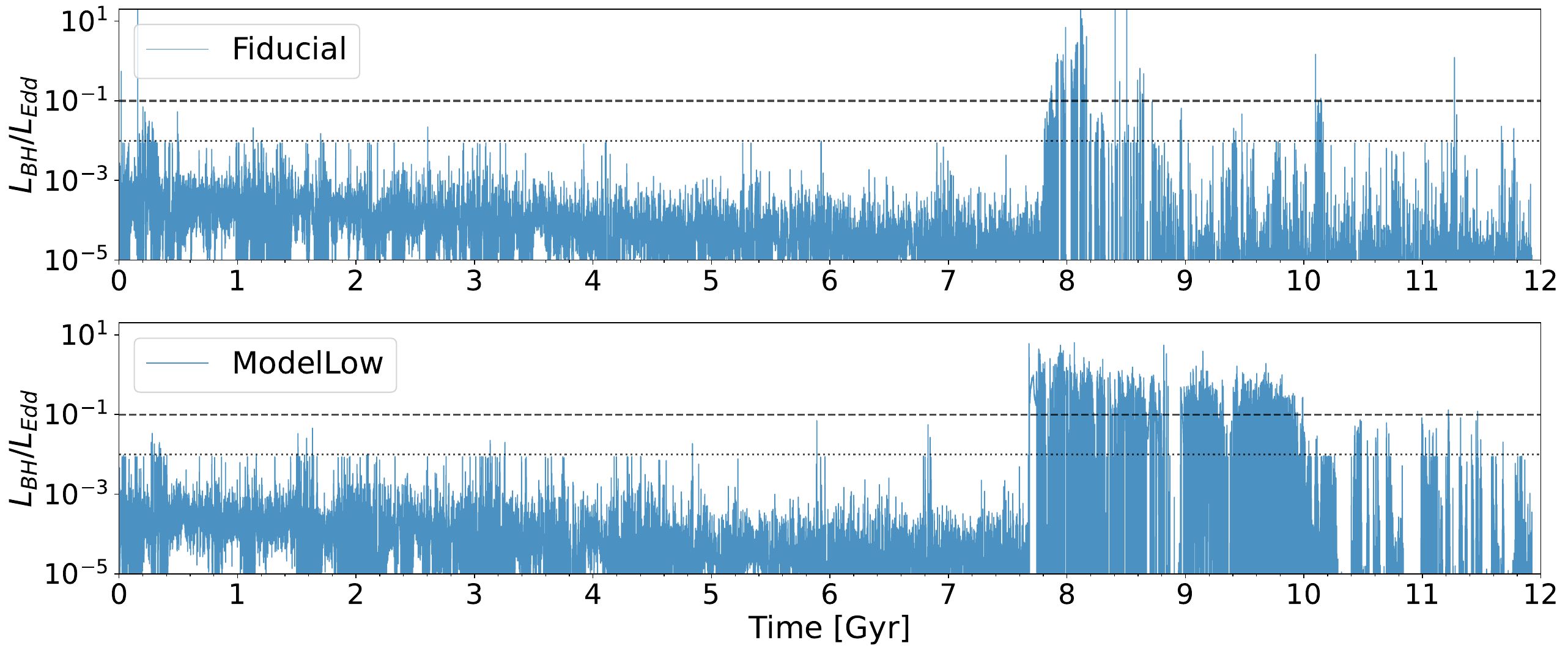}
\caption{Time evolution of the
AGN luminosity, normalized to the Eddington luminosity ($L/L_{\mathrm{Edd}}$), for the Fiducial run (top) and the ModelLow run (bottom). The dotted and dashed horizontal lines denote $0.01\,L_{\mathrm{Edd}}$ and $0.1\,L_{\mathrm{Edd}}$, respectively.
}
\label{fig:AGNlightcurve2}
\end{figure*}

Table~\ref{tab:sim-settings} summarizes the parameter settings for all models in our work. Our primary objective is to understand the role of AGN feedback in the evolution of disk galaxies. To address this comprehensively, we divide the suite into two categories—runs with AGN feedback and runs without—thereby enabling a direct comparison that isolates the impact of AGN feedback on galaxy evolution and quenching. In addition, we vary the value of \( \nu \) across the simulations; different choices represent distinct mechanisms or differing efficiencies of angular-momentum transfer. The third ``parameter'' we consider is the angular momentum of cold-mode cosmological inflows. In the high--angular-momentum case, we assume that the inflowing gas rotates at approximately the local Keplerian speed; in a contrasting case, we set the angular momentum of the cold-mode inflow to zero.

\subsection{AGN lightcurve}
\label{sec:AGNlightcurve}

Figure~\ref{fig:AGNlightcurve} presents the time evolution of the AGN luminosity, normalized to the Eddington luminosity \(\left(L/L_{\rm Edd}\right)\), for  four models. From top to bottom, the panels show Model0.25 \(\left(\nu=0.25\right)\), the Fiducial model \(\left(\nu=0.55\right)\), Model0.75 \(\left(\nu=0.75\right)\), and Model1.1 \(\left(\nu=1.1\right)\). The time-averaged luminosities for these runs are listed in Table~\ref{tab:agn-activity}. Up to \(t \sim 6\,\mathrm{Gyr}\), the luminosity histories exhibit no significant divergence among the four models, indicating that variations in \(\nu\) exert only a minor direct influence on AGN luminosity over this interval. Consistently, the time-averaged luminosities are also very similar (Table~\ref{tab:agn-activity}).

The parameter \(\nu\) governs the radial inflow velocity within the accretion disk. In principle, a larger \(\nu\) corresponds to more efficient angular-momentum transport and thus to a higher inflow rate (see Figure~\ref{fig:gas_inflow}), which would naively yield a higher AGN luminosity. Why, then, are the luminosities so similar across models? We attribute this to the self-regulating nature of AGN feedback, which mitigates the sensitivity to \(\nu\): when the inflow rate increases, the AGN brightens, and the resulting feedback interacts with the surrounding gas to suppress the accretion rate. In addition to this ``self-regulation'' effect, a larger \(\nu\) also enhances viscous dissipation, raising the gas temperature and lowering the gas density, which further reduces the inflow rate and damps the dependence of AGN luminosity on \(\nu\).

In the Fiducial model, we observe a pronounced increase in AGN luminosity at \(t \approx 7.8\,\mathrm{Gyr}\). Analysis of the simulation data indicates that this surge is driven by the accretion of substantial cold gas filaments that condense in the CGM. These filaments flow inward, merge together, and ultimately accrete onto the central supermassive black hole, triggering AGN activity and boosting its luminosity. The formation of these cold filaments arises from radiative cooling of CGM gas; further details are provided in Section~\ref{filamentformation}.

\begin{figure*}
\centering
\includegraphics[width=0.8\textwidth]{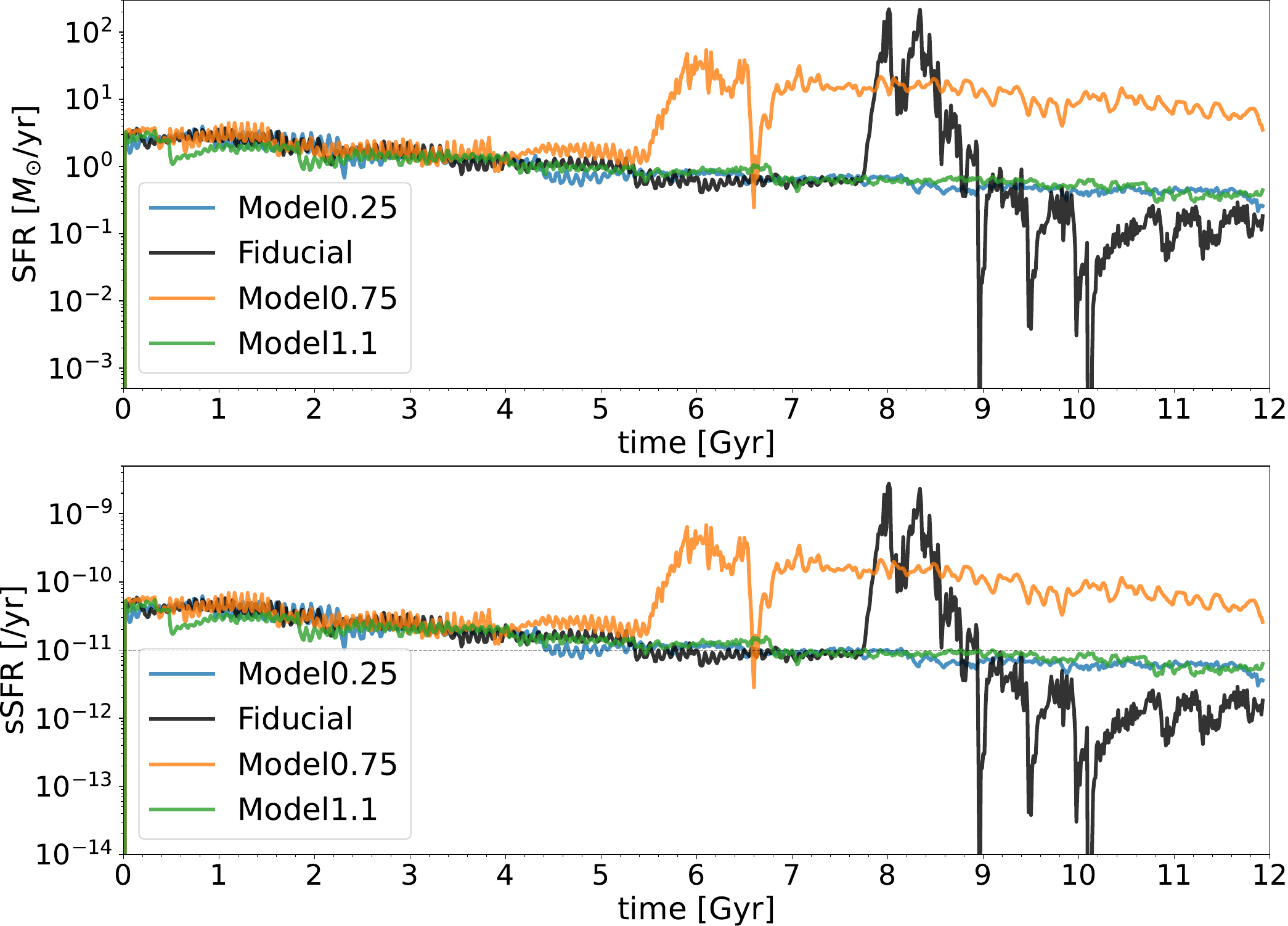}
\caption{Time evolution of the star formation rate (SFR, top) and specific star formation rate (sSFR, bottom) in various models. The dotted horizontal line marks the quenching threshold; values below this line indicate that the galaxy is quenched. The peak at $\sim $ 8Gyr in the Fiducial model arises from the same physical origin with the peak of AGN luminosity shown in Fig. \ref{fig:AGNlightcurve}, i.e., the formation of cold filaments in the CGM and their subsequent infall onto the galaxy. The galaxy in the Fiducial model is quenched within $\sim 1$ Gyr by the AGN feedback.
}
\label{fig:SFRcurve}
\end{figure*}

In contrast to the Fiducial model, the other three runs with different viscosity parameters \((\nu)\) do not exhibit a comparably strong increase in AGN luminosity. In Model0.25 and Model1.1, no cold filaments form throughout the simulation. For Model0.75 (third row from the top in Figure~\ref{fig:AGNlightcurve}), inspection of the time-resolved simulation visualization reveals the formation of substantial cold gas filaments in the CGM as in the Fiducial model. However, the gas in these filaments are predominantly consumed by star formation (see the orange curve in Figure~\ref{fig:SFRcurve}), leaving only a small fraction to accrete onto the black hole, as reflected by the modest feature in the AGN light curve at \(t \approx 5.8\,\mathrm{Gyr}\). This naturally raises two questions: why do no cold filaments form in Model0.25 and Model1.1? And why, although both the Fiducial model and Model0.75 form cold filaments in the CGM, does only the Fiducial model trigger a luminous AGN, whereas Model0.75 primarily boosts the star formation rate (and for an extended period)? We address these questions in Section~\ref{filamentformation}.

Figure~\ref{fig:AGNlightcurve2} illustrates the time evolution of the AGN luminosity, normalized to the Eddington luminosity \(\left(L/L_{\rm Edd}\right)\), for the ModelLow run; for comparison, the light curve of the Fiducial model is also shown. Prior to \(\sim 8\,\mathrm{Gyr}\), the luminosity of ModelLow is slightly higher than that of the Fiducial model because the cold-mode cosmological inflow in ModelLow has lower angular momentum and thus more readily accretes onto the black hole. After \(t \gtrsim 7.8\,\mathrm{Gyr}\), the AGN luminosity in ModelLow is generally significantly higher, and the high-luminosity phase lasts much longer than in the Fiducial model. The reason is as follows. In both cases, the high luminosity is due to the accretion of cold filaments, originally formed in the CGM, by the supermassive black hole. Around \(8\,\mathrm{Gyr}\), the AGN in the Fiducial model reaches higher power and therefore produces stronger winds; these winds push the cold filaments away from the black hole on a shorter timescale.

\subsection{AGN duty cycle}

In addition to the time-averaged AGN luminosities for the five models, Table~\ref{tab:agn-activity} also reports the AGN duty cycle and the time-averaged star formation rate (SFR). According to \citet{Greene_2007}, the duty cycle for black holes with masses around \(10^{7}\,M_{\odot}\) is approximately \(0.4\%\). In our simulations, the initial black hole mass was set to \(5\times 10^{7}\,M_{\odot}\). In the Fiducial model, the AGN duty cycle is \(0.49\%\), closely aligning with observational constraints and indicating moderate activity levels. 

By contrast, all other models fail to reproduce the observed AGN duty cycle.
Specifically, reducing the angular momentum of the injected cold filaments leads to a sharp increase in the duty cycle, reaching \(7.10\%\) in ModelLow. This value is approximately an order of magnitude higher than the observational estimate. The primary cause is that, as discussed at the end of \S\ref{sec:AGNlightcurve}, the cold filaments in ModelLow are more resistant to expulsion by the central AGN than in the Fiducial model.
\subsection{SFR and the quenching of galaxy}

\begin{figure*}
\centering
\includegraphics[width=\textwidth]{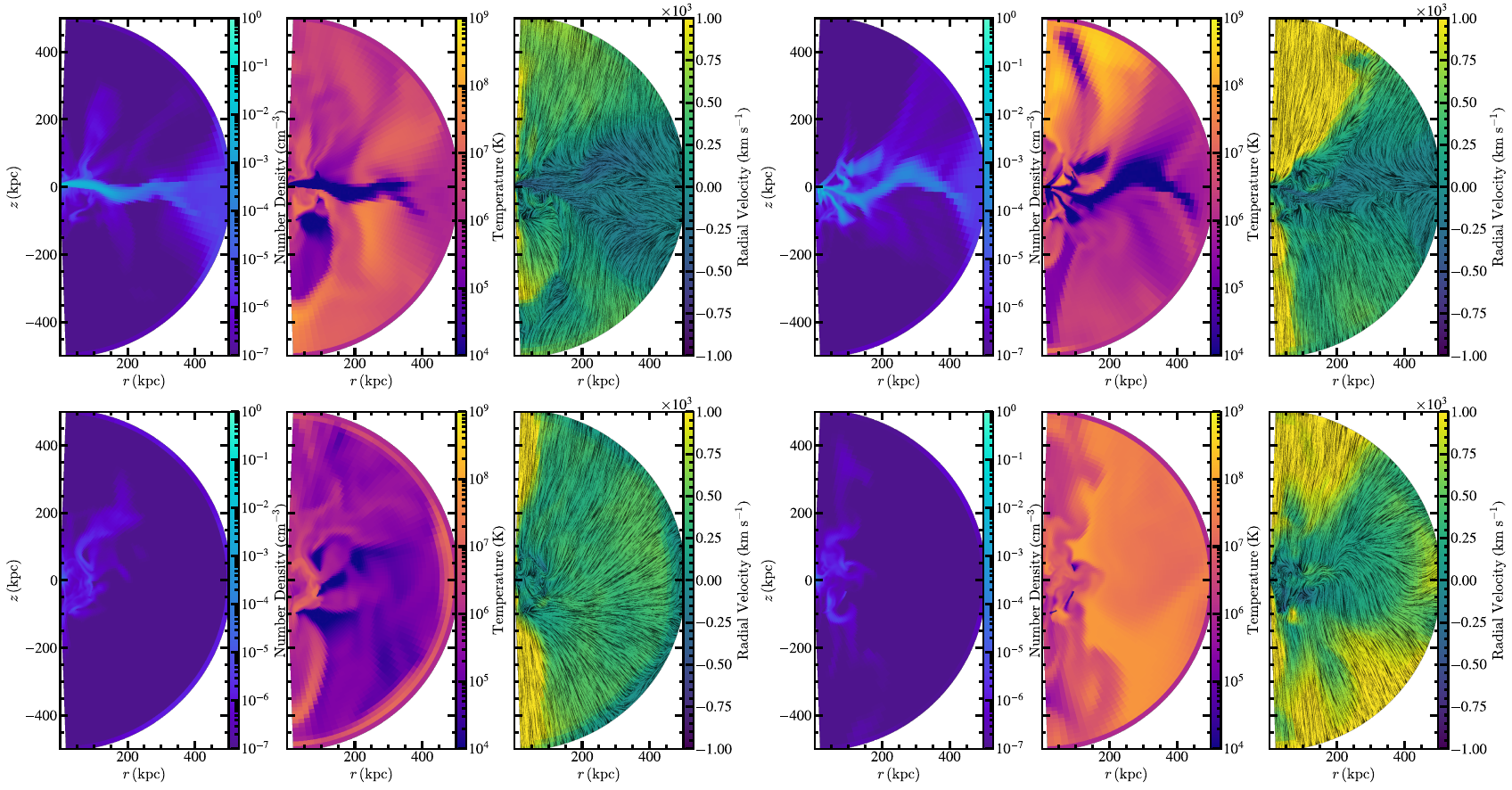}
\caption{Snapshots of the spatial distributions of number density, temperature, and radial velocity, taken at $t = 7.7625~\mathrm{Gyr}$ for the Fiducial (top left), Model0.75 (top right), Model0.25 (bottom left), and Model1.1 (bottom right) models. In the rightmost panel, black streamlines are overlaid on the panel; their local tangents indicate the direction of the velocity field. In the upper two models, cold filaments have condensed and coalesced and are migrating inward preferentially along the equatorial plane, where they will eventually join the ISM. An animation showing the time evolution of these models is available at \href{https://zenodo.org/records/18397144}{https://zenodo.org/records/18397144}.}
\label{fig:coldfilament}
\end{figure*}

\begin{figure*}
\centering
\includegraphics[width=\textwidth]{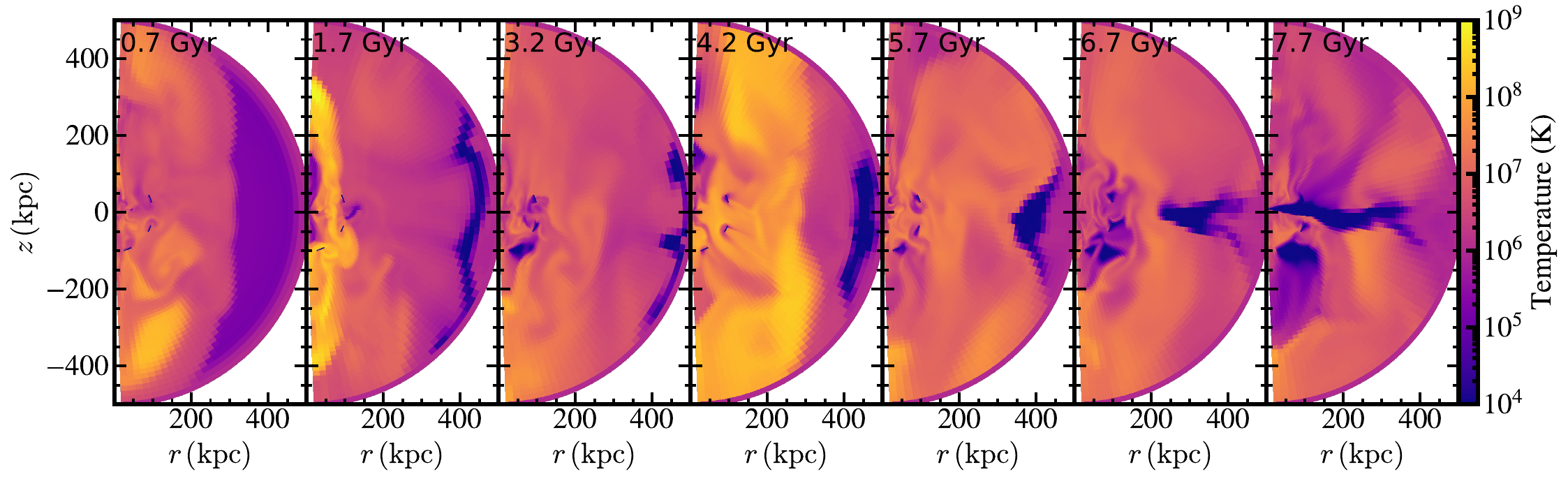}
\caption{Time sequence of the distribution of gas temperature maps in the Fiducial model, showing the AGN-driven formation of cold filaments in the CGM and their equatorial infall into the center of the galaxy. Temperatures are given in Kelvin, with a common color bar on the far right. }
\label{fig:coldfilament_formation}
\end{figure*}

\begin{figure}
\centering
\includegraphics[width=0.9\columnwidth]{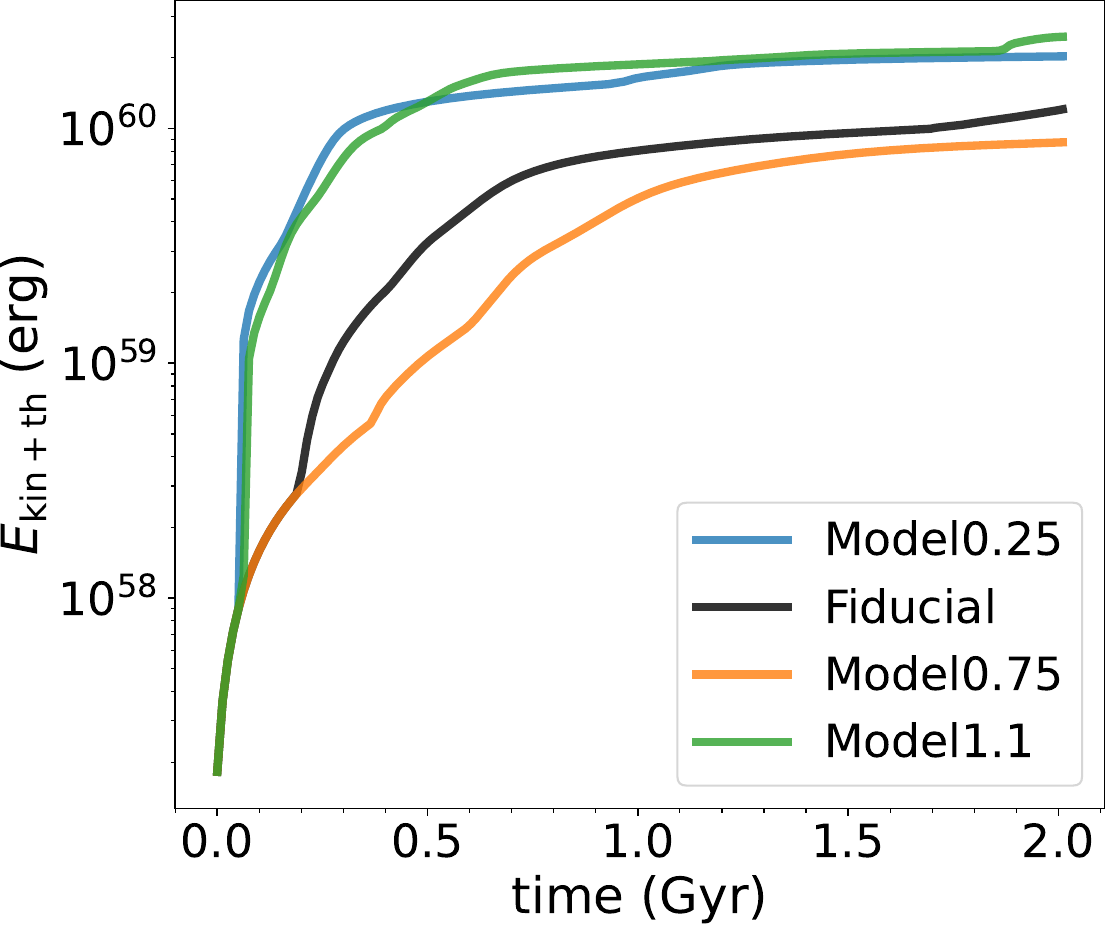} 
\caption{Cumulative kinetic plus thermal energy of the gas that has crossed the \(300\,\mathrm{kpc}\) radius within the first \(2\,\mathrm{Gyr}\) of the simulation for different models. The values in the Fiducial and Model0.75 runs are significantly lower than those in the other two models, explaining why cold filaments form only in these two models.}
\label{fig:cumulative_energy}
\end{figure}

\begin{figure*}
\centering
\includegraphics[width=0.8\textwidth]{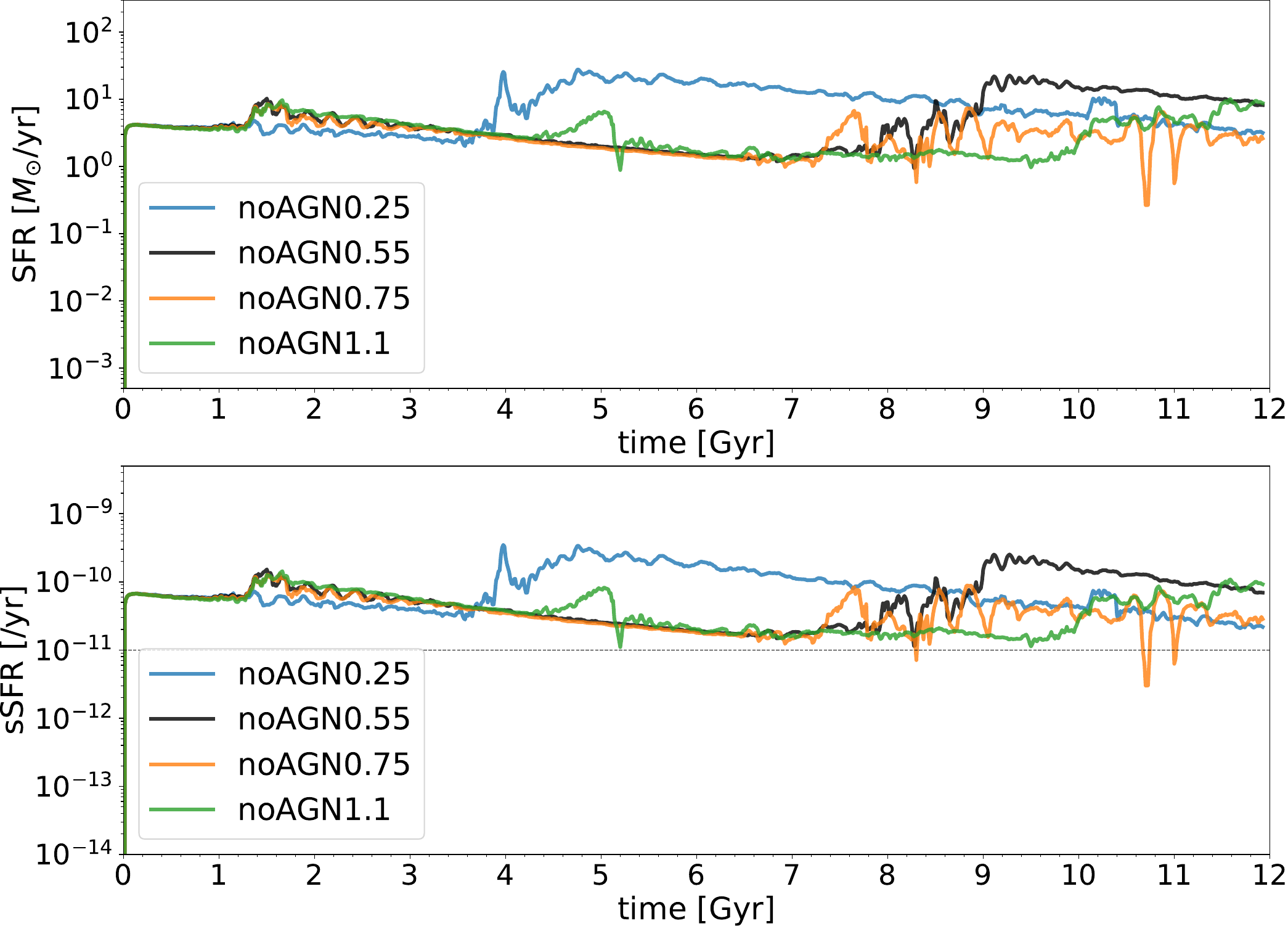}
\caption{Time evolution of the star formation rate (SFR, top) and specific star formation rate (sSFR, bottom)  with AGN feedback disabled. The dotted horizontal line marks the quenching threshold; values below this line indicate that the galaxy is quenched. No galaxies are quenched without AGN feedback. The peak SFR in noAGN0.55 run is significantly lower than in the Fiducial model shown in Fig. \ref{fig:SFRcurve}. This indicates that AGN feedback facilitates the formation of more massive cold filaments, thereby promoting intense starburst episodes.
}
\label{fig:SFRcurve_noAGN}
\end{figure*}

\begin{figure*}
\centering
\includegraphics[width=0.8\textwidth]{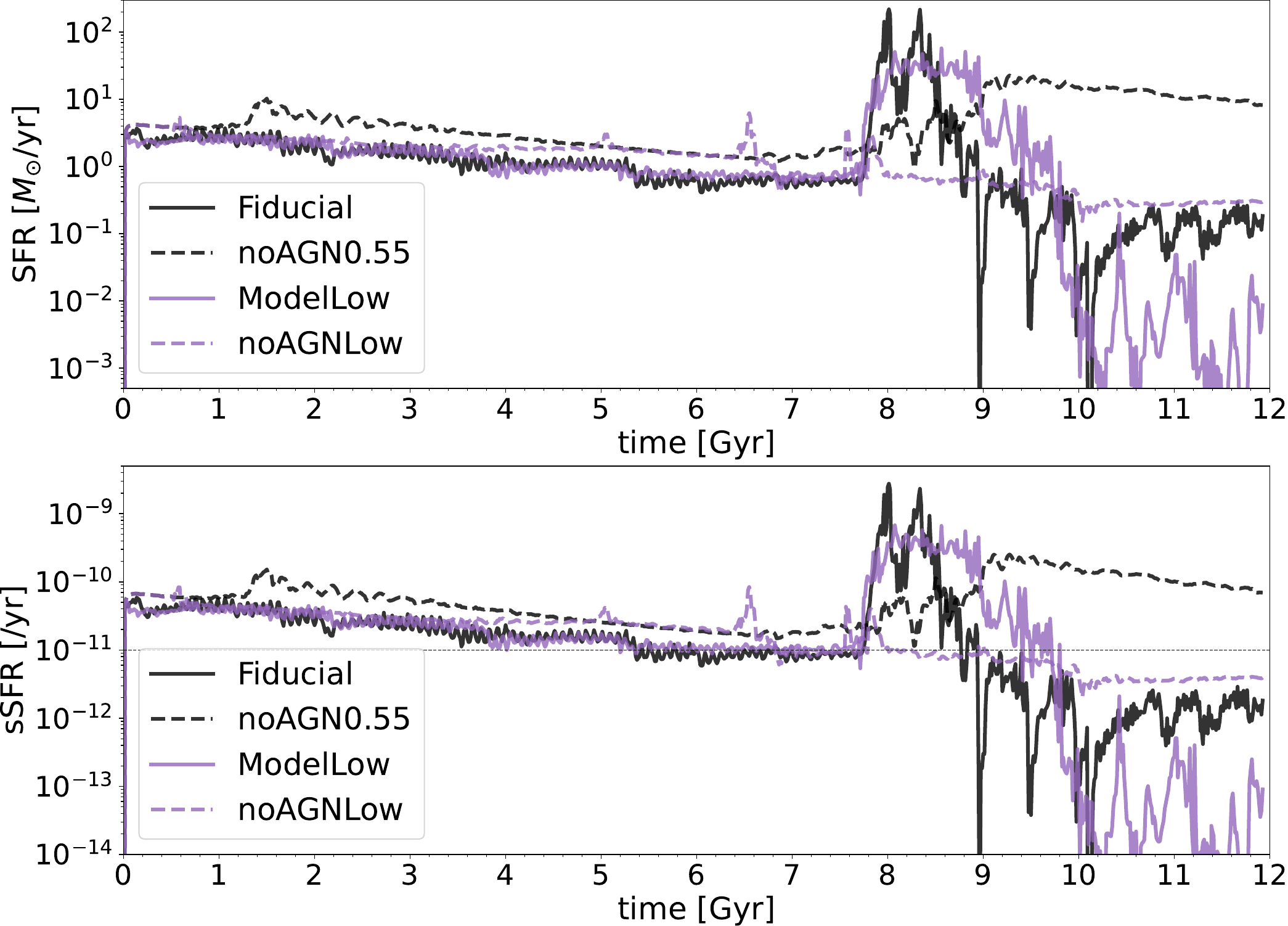}
\caption{
Time evolution of the SFR (top) and sSFR (bottom) for different models. A similar quenching pattern is observed in both the Fiducial and ModelLow runs.
The pronounced rise in SFR after $\sim 8$ Gyr in the noAGN0.55 run results from the formation and subsequent infall of cold filaments, while its absence in the noAGNLow run is likely caused by the presence of a more massive black hole. }
\label{fig:SFRcurve_Low}
\end{figure*}

The SFR and specific star formation rate (sSFR) for the entire galaxy across several models are shown in Figure~\ref{fig:SFRcurve}. The dotted horizontal line in the lower panel marks the quenching threshold. Note that this threshold applies only at low redshift; at higher redshift it would be substantially higher. Because our initial conditions are designed to represent a disk galaxy in the local Universe rather than a typical \(z\sim2\text{--}3\) system, we adopt this threshold value for illustrative (schematic) purposes. From the figure, we see that in the Fiducial model the SFR rises rapidly around \(t \approx 7.5\,\mathrm{Gyr}\), then declines quickly near \(t \sim 8.5\,\mathrm{Gyr}\), and the galaxy is quenched by \(t \sim 9\,\mathrm{Gyr}\). It is interesting to note that, after the quenching, the galaxy transitions to a dynamically hot state with a large stellar bulge formed. The detailed physical mechanism responsible for quenching and the comparison of the bulge-to-disk ratio between predictions and observations will be discussed in a subsequent paper in this series. As in the Fiducial model, the SFR in Model0.75 also increases significantly (although earlier in time). In contrast, in Model0.25 and Model1.1 the SFR gradually declines over the course of the evolution. We find that the increase in SFR observed in both the Fiducial model and Model0.75 is due to the infall of cold filaments from the CGM into the galaxy. We will discuss the physical mechanism of filaments formation in \S\ref{filamentformation}.

However, significant differences exist between the Fiducial model and Model0.75. First, in Model0.75, although these filaments penetrate deeply into the galaxy’s core—as indicated by the rapid rise in SFR at \(t \sim 6\,\mathrm{Gyr}\)—they largely fail to reach the central SMBH. As shown in Figure~\ref{fig:AGNlightcurve}, there are only subtle signatures of enhanced AGN luminosity between \(t \sim 6\)–\(7\,\mathrm{Gyr}\). Second, in Model0.75, the SFR (and sSFR) remains high from \(t \sim 6\,\mathrm{Gyr}\) through the end of our simulation, whereas in the Fiducial model, the SFR declines sharply after \(t \sim 8.5\,\mathrm{Gyr}\) and the galaxy is quenched. These two differences are, in fact, closely related. In the Fiducial model, AGN activity is significantly enhanced, driving powerful feedback and suppressing SFR, which ultimately quenches the galaxy. By contrast, in Model0.75 the AGN activity is only modestly enhanced and thus insufficient to quench the galaxy, allowing the SFR to remain high. The reason for the differences will be discussed in \S\ref{filamentformation}.

In Fig.~\ref{fig:SFRcurve}, the galaxy-wide SFR exhibits oscillations prior to \(t \approx 7\,\mathrm{Gyr}\). These oscillations are driven by AGN feedback processes that excite approximately simple-harmonic variations of the disk gas mass, most prominently at radii \(R \sim 4\text{--}6\,\mathrm{kpc}\). Consistent with this interpretation, the no-AGN control run shows no such oscillations. The characteristic period, \(P \simeq 0.1\,\mathrm{Gyr}\), is not tied to the AGN light-curve variability; rather, it is comparable to the local dynamical time. This picture accords with the findings of \citet{Peng_2025}, who argue that potential perturbations induced by AGN outflows—together with the injection of newly formed stars within those outflows and their subsequent dynamical friction—can drive bulge oscillations.

\subsection{Correlation between SFR and BHAR}

Combining Figures~\ref{fig:AGNlightcurve} and \ref{fig:SFRcurve}, we find a clear positive correlation between the SFR and the AGN luminosity (equivalently, the black hole accretion rate; BHAR). Taking the Fiducial model as an example, just before \(t \sim 8\,\mathrm{Gyr}\), both quantities rise rapidly, whereas after \(t \sim 9\,\mathrm{Gyr}\), both decline quickly. Our simulations clearly show that this concurrence arises because both the luminous AGN episodes and the enhanced SFR are fueled by the same gas reservoir. Moreover, a detailed time cross-correlation analysis indicates that the AGN luminosity lags the SFR by a finite interval. A comprehensive analysis of the BHAR-SFR correlation, its comparison with observations, and the time lag between SFR and BHAR will be presented in a subsequent paper in this series. We find that both the slope and the normalization of the correlation are in good agreement with observations.

Based on the Fiducial model results, we emphasize that, with a physical model of AGN feedback, both the positive BHAR–SFR correlation and AGN-driven quenching can be reproduced simultaneously and self-consistently. In other words, the presence of a positive correlation does not argue against AGN feedback as the galaxy quenching mechanism.

\subsection{Formation of cold filaments}
\label{filamentformation}

As noted above, we find that the rapid increase in SFR in the Fiducial model and Model0.75 arises from the formation of cold filaments in the CGM and their infall toward the central regions of the galaxy. To illustrate this, Figure~\ref{fig:coldfilament} presents contemporaneous snapshots of the four models. From left to right, the panels show the two-dimensional distributions of the number density, temperature, and the abundance of the hot-mode cosmological inflow. In the two models in the top row (i.e., Fiducial and Model0.75), cold filaments coalesce into a cold stream and fall toward the galactic center approximately along the equatorial plane. 
By contrast, in Model0.25 and Model1.1, no cold filaments form. Correspondingly, neither the AGN luminosity nor the SFR exhibits any significant enhancement (see Figures~\ref{fig:AGNlightcurve} and \ref{fig:SFRcurve}). What is the physical origin of the differences among the four models? Why do cold filaments form in the Fiducial and Model0.75 models but not in the others?

To answer these questions, we first need to understand the physical mechanism for the formation of the cold filaments in the CGM. For this purpose, we draw Figure \ref{fig:coldfilament_formation}, which comprises seven horizontally arranged panels, each showing the two-dimensional distribution of gas temperature at different times in the Fiducial model. From left to right, recurrent AGN feedback from episodic AGN activities expels gas from the galaxy into the CGM. The gas cools radiatively and condenses into cold filaments in the CGM. These filaments are found to gradually collapse toward the equatorial plane and stream inward along the equatorial plane, ultimately accreting onto the center of the galaxy.

Consistent with this physical picture, our results indicate that the gas accumulated in the CGM's cold filaments is dominated by material originating in the ISM in the galaxy's central region rather than by in-situ cooling of the ambient CGM; moreover, tracer analysis constrains the contribution from cosmological hot-mode inflow within these filaments to at most $20\%$ throughout the simulation. The contribution from cosmological cold-mode inflow within these filaments is negligible, with a tracer-inferred mass fraction $\lesssim 10^{-3}$ at all times. The mass budget supports this conclusion: the initial CGM in our setup contains only $2\times 10^{10}\,M_\odot$ of gas, whereas the cold filaments eventually reach $\sim1.2\times 10^{11}\,M_\odot$. Thus, the filaments' mass cannot be supplied by the original CGM alone. The filaments are not formed solely through a single episode of local CGM cooling, nor are they dominated by cosmological hot-mode inflow. Instead, repeated AGN-feedback-induced shocks expel ISM gas into the CGM, where it cools and condenses into filaments.

Because the galaxy initial conditions and the cosmological inflow are identical across the four models, we propose that cold-filament formation is regulated by the net level of feedback heating in the CGM. If this heating is sufficiently strong, CGM gas cannot cool efficiently and therefore fails to condense into cold filaments. Previous work suggests that such heating is often dominated by the energy deposited by AGN-driven winds into the CGM \citep{2023MNRAS.524.5787Z}; here, however, our focus is on the total feedback energy that reaches the CGM, summed over all channels.

To quantify this, we computed the cumulative energy transported into the CGM for the four models. This energy budget includes contributions from all feedback channels, including both AGN and stellar feedback, and accounts for both kinetic and thermal energy. Figure~\ref{fig:cumulative_energy} shows the cumulative energy crossing a spherical surface at \(r=300\,\mathrm{kpc}\) during the first \(2\,\mathrm{Gyr}\) of evolution, during which cold filaments form in the Fiducial and Model0.75 runs. We find that the total energy transported to \(r=300\,\mathrm{kpc}\) in Model0.25 and Model1.1 is consistently a factor of \(3\text{--}4\) larger than in the Fiducial and Model0.75 models. This higher CGM energy transport effectively suppresses radiative cooling, which explains the absence of cold-filament formation in Model0.25 and Model1.1.

We find that both the AGN power—which sets the wind power \citepalias{2018ApJ...857..121Y}—and the SFR (and thus the strength of stellar feedback) show little variation among the four models during the first \(2\,\mathrm{Gyr}\). We therefore argue that the differences in the cumulative kinetic energy in Figure~\ref{fig:cumulative_energy} are more likely driven by differences in the gas density within the galaxy. We have computed the radial profiles of the \(\theta\)-averaged ISM density for all four models and find that the Fiducial and Model0.75 runs have the highest densities over most galactocentric radii. Consequently, in these two models a larger fraction of the AGN energy is deposited within the ISM, leaving less energy to propagate out to \(r=300\,\mathrm{kpc}\).Why is the gas density highest in these two models? This is a complex question, because the ISM density is set by the coupled action of multiple processes, including the radial inflow velocity (related to the parameter \(\nu\)) as well as AGN and supernova feedback. Finally, while the Fiducial model is arguably the most plausible representation of real galaxies—given that it reproduces the observed AGN duty cycle—it remains unclear whether the other three models have equally realistic counterparts in nature.

Another question is why the cold filaments in Model0.75 are largely consumed before they can fuel the black hole. The mass of the cold filaments formed in the two models are comparable, which is  \(1.23\times10^{11}\,M_\odot\) in the Fiducial model and  \(1.64\times10^{11}\,M_\odot\) in Model0.75. The filaments in Model0.75 are more massive because the cumulative total energy in Model0.75 is smaller, as shown in Figure~\ref{fig:cumulative_energy}. The small difference in the filament mass implies that the discrepancy of mass should not be the reason. We find that when the filaments in Model0.75 approach to within $\sim$ 10 kpc, the AGN is very strong, with $L\sim 2\%L_{\rm Edd}$. The strong wind launched from the AGN disrupts the cold filaments, preventing them from directly falling onto the black hole. Instead, the gas is redistributed into an extended disk near the equatorial plane, which connects back to the original galactic gaseous disk. As a result, the black hole accretion rate does not experience the strong enhancement seen in the Fiducial model.  Conversely, this process leaves a substantial reservoir of cold gas in the disk, sustaining the prolonged high SFR observed in Model0.75 (Figure~\ref{fig:SFRcurve}). We note that while galaxies corresponding to Model0.75 may be less common than the Fiducial case (since most of the time the AGN is not luminous), observed bulgeless giant disk galaxies \citep{2010ApJ...723...54K} may represent its real-world counterparts. A more detailed discussion of this scenario will be presented in a subsequent paper.

\subsection{Comparison with the cases without AGN feedback and with low-angular-momentum cold inflow}

To assess the role of AGN feedback in regulating the SFR, we conducted simulations of five models without AGN feedback, namely ``noAGN0.25'', ``noAGN0.55'', ``noAGN0.75'', ``noAGN1.1'', and ``noAGNLow'', as summarized in Table~\ref{tab:sim-settings}. Figure~\ref{fig:SFRcurve_noAGN} shows the evolution of the SFR for the first four models. Compared to the models with AGN feedback, these four runs exhibit SFR values that are several times higher and remain relatively stable over time; that is, the galaxy is not quenched. These results are expected. However, it is noteworthy that without AGN feedback the SFR does not reach exceptionally high values, peaking at only \(\sim 20\,M_\odot\,\mathrm{yr}^{-1}\). By contrast, the SFR in the Fiducial model and in Model0.75 can reach or exceed \(100\,M_\odot\,\mathrm{yr}^{-1}\), occasionally approaching \(200\,M_\odot\,\mathrm{yr}^{-1}\). 

A key question then arises: why does the inclusion of AGN feedback lead to a higher peak SFR? As we have analyzed in \S\ref{filamentformation}, we find that when AGN feedback is present, the AGN wind expels galactic gas into the CGM, leading to stronger gas accumulation in that region over long timescale of evolution. As a result, the cold filaments that forms under these conditions is significantly more massive. When the large filament eventually falls back onto the galaxy, it triggers a more intense star formation episode. In this way, AGN feedback plays an accumulatively positive role in promoting the high-SFR star-burst phase.  

In the Fiducial model, the angular momentum of the injected cold-mode cosmological inflow is roughly equal to the local Keplerian value. To examine the effect of angular momentum, we also simulated a ``ModelLow'' run in which the angular momentum of the cold inflow is zero (Table~\ref{tab:sim-settings}). For comparison, we additionally simulated a model with low-angular-momentum cold inflow but without AGN feedback (``noAGNLow''). Figure~\ref{fig:SFRcurve_Low} shows the evolution of the SFR for ``ModelLow'' and ``noAGNLow'', together with the Fiducial and ``noAGN0.55'' models. The SFR history of ModelLow resembles that of the Fiducial model in that both exhibit a high peak around \(t \sim 8\,\mathrm{Gyr}\), followed by quenching. The similarity arises because, in both cases, the peak is produced by the infall of cold filaments formed in the CGM via radiative cooling. However, the SFR outburst in ModelLow persists for a longer period than in the Fiducial model. This is consistent with Figure~\ref{fig:AGNlightcurve2}, which shows that the high-luminosity phase in ModelLow lasts longer than in the Fiducial model. The reason has been provided at the end of \S\ref{sec:AGNlightcurve}.

Interestingly, the galaxy in the noAGNLow model is also quenched. The SFR in this model gradually declines, likely owing to the aging of the stellar population: as stars age, they produce fewer stellar winds, resulting in progressively less recycled gas to fuel star formation. In contrast, the galaxy in the noAGN0.55 model is never quenched; in fact, the SFR even increases after \(t \sim 8\,\mathrm{Gyr}\), which we attribute to the formation of cold filaments in the CGM. The difference between noAGN0.55 and noAGNLow is that the angular momentum of the cold inflow in the noAGN0.55 model is high. Consequently, the infall speed of the cold cosmological inflow is much higher in the noAGNLow model, and the black hole mass in this model is significantly larger --- the final black hole mass in this model is about ten times higher than in the noAGN0.55 model, \(\sim 10^{10}\,M_\odot\). The heavier black hole produces a much deeper gravitational potential, thereby increasing the infall speed and lowering the gas density. This explains why the SFR of noAGNLow is lower than that of noAGN0.55 throughout the evolution. The influence of the heavier black hole even extends into the CGM, although gravity there is not dominated by the black hole. Because the gravitational potential is deeper in the noAGNLow model, the gas pressure–gradient force within the galaxy is smaller, and thus the CGM density is lower than in the noAGN0.55 model. The lower density is the reason why cold filaments do not form in noAGNLow model.

\section{Summary and Discussion}
\label{sec:Discussion}

In our previous works, we have constructed the MACER framework to investigate the evolution of a single galaxy, focusing on the role of AGN feedback \citep{2018ApJ...857..121Y,2025ApJ...985..178Z}. 

In our previous studies, we used MACER to systematically investigate the evolution of elliptical galaxies. In the present work and in several subsequent papers, we focus on AGN feedback in disk galaxies. This paper presents the model setup and the general results. Follow-up papers will examine in detail the physical mechanism of galaxy quenching, the correlation and time lag between BHAR and SFR, the radial profile of the CGM X-ray surface brightness, the (cold) gas fraction in galaxies across a range of AGN luminosities, and other topics. Because our simulations are two-dimensional, we introduce a phenomenological description of ``viscous stress'' to approximate angular-momentum transport in the galaxy. We explore a range of viscosity parameters to identify values that plausibly mimic real systems. On the one hand, the self-regulating nature of the system implies that the precise value of the viscosity is not critical; on the other hand, different parameter choices do lead to distinct predictions—for example, regarding the formation of cold filaments and the AGN duty cycle.

The main results of the Fiducial model, which best reproduces the observations, are summarized below.

\begin{itemize}
\item \textit{AGN light curve and duty cycle.} The AGN light curve predicted by the Fiducial model is shown in the second panel of Figure~\ref{fig:AGNlightcurve}. The AGN is not very luminous most of the time, but it occasionally reaches a quasar phase with a high Eddington ratio, and sometimes even becomes super-Eddington. The predicted AGN duty cycle is \(0.49\%\), consistent with the observational result reported by \citet{Greene_2007}.

\item \textit{SFR and the quenching of the galaxy.} The time evolution of the SFR is presented in Figure~\ref{fig:SFRcurve}. As shown in the figure, the SFR sometimes reaches values as high as \(>\!100\,M_\odot\,\mathrm{yr}^{-1}\). This starburst phase lasts for nearly \(1\,\mathrm{Gyr}\). After the starburst phase, the SFR rapidly decreases to very low values, and the galaxy becomes quenched. The quenching is caused by AGN feedback, and the detailed analysis will be presented in a subsequent paper.

\item \textit{Correlation between BHAR and SFR.} By comparing Figures~\ref{fig:AGNlightcurve} and \ref{fig:SFRcurve}, we see that the BHAR and SFR are positively correlated. Specifically, the concurrent enhancements in SFR and AGN luminosity are caused by the infall of cold filaments formed in the CGM through radiative cooling of the gas. These cold filaments fuel both star formation and the growth of the supermassive black hole.
So the quenching of the galaxy by AGN feedback  and the positive correlation between BHAR and SFR are reproduced simultaneously in our Fiducial model. This indicates that the presence of a positive correlation does not argue against AGN feedback as the galaxy quenching mechanism.

\item \textit{Accumulative AGN feedback facilitates the presence of high-SFR starburst episodes.} Comparing the SFR without AGN feedback (Fig. \ref{fig:SFRcurve_noAGN}) with those with AGN feedback (Fig. \ref{fig:SFRcurve}), we find that the peak value of SFR in the former is about 10 times  lower than in the latter. This is because AGN winds expel gas into the CGM, leading to stronger gas accumulation and formation of more massive cold filaments, which triggers more intense star formation. 

\end{itemize}

There are several caveats in our work that could be addressed in future studies.
\begin{itemize}
    \item Our simulation is two-dimensional. While this 2D framework captures the key dynamical processes, it neglects a few three-dimensional effects which might quantitatively alter some of our results, such as the cascade of turbulence. In addition, the evolution of hydrodynamic and thermal instabilities also slightly differ in 2D due to the inverse turbulent cascade. Investigation on the evolution of disk galaxy under our 3D framework \citep{2025ApJ...985..178Z} will be carried out in future.
    
    \item  Our simulations remain somewhat idealized: we do not adopt a more realistic high-\(z\) density; several important physical processes such as galaxy mergers and the evolution of the galactic gravitational potential are not incorporated. Our goal in this series is to isolate and investigate the key physical mechanisms that regulate galaxy evolution, rather than to achieve a precise quantitative agreement with observations. We plan to address these limitations by performing cosmological zoom-in simulations. 
In the future, we plan to refine our model by addressing these issues step by step. As the simulations become progressively more physical and realistic, we will compare the results with observations across various aspects, such as the SFR, CGM filaments kinematics, and galactic winds.
\end{itemize}

\begin{acknowledgments}
{\it Acknowledgments} Y.Z., F.Y., and S.J. are supported by the NSF of China (grants 12192220, 12192223, 12522301, 12133008, and 12361161601), the China Manned Space Program (grants CMS-CSST-2025-A08 and CMS-CSST-2025-A10), and the National Key R\&D Program of China No. 2023YFB3002502, {and the National SKA Program of China (No. 2025SKA0130100)}. Numerical calculations were run on the CFFF platform of Fudan University, the supercomputing system at the Supercomputing Center of Wuhan University, and the High Performance Computing Resource at the Core Facility for Advanced Research Computing at Shanghai Astronomical Observatory. 
\end{acknowledgments}

\noindent{\it Data Availability} The data underlying all figures and tables in this paper (i.e., the reduced/derived data products used to generate the plots and tables) are available from the corresponding authors upon reasonable request. The full simulation outputs (e.g., snapshots and related files) are also available upon request. Data will be shared by the corresponding authors via file transfer or a download link. Requests should specify the relevant run(s), the required time range/snapshots, and the variables/data products needed. We welcome scientific collaboration. Corresponding authors: Feng Yuan (\href{mailto:fyuan@fudan.edu.cn}{fyuan@fudan.edu.cn}) and Suoqing Ji (\href{mailto:sqji@fudan.edu.cn}{sqji@fudan.edu.cn}).

\bibliographystyle{aasjournal}
\bibliography{sample631}

\end{document}